\RequirePackage{accents}
\documentclass[epjH]{svjour}
\def\makeheadbox{\relax}
\usepackage{graphicx,stackengine}
\usepackage[space]{cite}
\usepackage[caption = false]{subfig}
\usepackage{float}
\usepackage{xcolor} 
\usepackage{amssymb}
\usepackage{verbatim}
\begin{document}
\newpage
\newcommand{\sstw}{\ensuremath{\sin^2 \theta_{\mathrm W} }}
\newcommand{\sqtw}{\ensuremath{\sin^4 \theta_{\mathrm W} }}
\newcommand{\cstw}{\ensuremath{\cos^2 \theta_{\mathrm W} }}
\newcommand{\stw}{\ensuremath{\sin \theta_{\mathrm W} }}
\newcommand{\ctw}{\ensuremath{\cos \theta_{\mathrm W} }}
\newcommand{\ctcab}{\ensuremath{\cos \theta_{\mathrm C} }}
\newcommand{\stcab}{\ensuremath{\sin \theta_{\mathrm C} }}
\newcommand{\tw}{\ensuremath{\theta_{\mathrm W} }}
\newcommand{\Pe}{\mathrm e}
\newcommand{\Pp}{\mathrm p}
\newcommand{\Pn}{\mathrm n}
\newcommand{\PD}{\mathrm {D_2}}
\newcommand{\Pmup}{\mathrm {\mu^+}}
\newcommand{\Pmum}{\mathrm {\mu^-}}
\newcommand{\Pmupm}{\mathrm {\mu^\pm}}
\newcommand{\Pep}{\mathrm {e^+}}
\newcommand{\Pem}{\mathrm {e^-}}
\newcommand{\PK}{\mathrm{K}}
\newcommand{\Ppiz}{\pi^\circ}
\newcommand{\Ppi}{\pi}
\newcommand{\pbarp}{\mathrm{p}\bar{\mathrm{p}}}
\newcommand{\Pnu}{\nu}
\newcommand{\APnu}{{\bar{\nu}}}
\newcommand{\Pnue}{\nu_{\mathrm{e}}}
\newcommand{\APnue}{\bar{\nu}_{\mathrm{e}}}
\newcommand{\Pnul}{\nu_\ell}
\newcommand{\APnul}{\bar{\nu}_l}
\newcommand{\Pnumu}{\nu_\mu}
\newcommand{\Pnutau}{\nu_\tau}
\newcommand{\APnutau}{\bar{\nu}_\tau}
\newcommand{\APnumu}{\bar{\nu}_\mu}
\newcommand{\numupbarp}{{\stackon[+0.2pt]{$\nu_\mu$}{$\scriptscriptstyle\smash{({-})}\hphantom{\nu} $}}}
\newcommand{\nuepbarp}{{\stackon[+0.2pt]{$\nu_\mathrm{e}$}{$\scriptscriptstyle\smash{({-})}\hphantom{\nu} $}}}
\newcommand{\pM}{\mathrel{\raise -3pt \hbox{\tiny(}\!
                 \raise -3pt \hbox{+} 
                 \settowidth {\dimen03} {+}
                 \hskip-\dimen03
                 \raise -6.5pt \hbox {$-$}
                 \!\raise -3pt \hbox{\tiny)}}}
\newcommand{\PW}{\mathrm W}
\newcommand{\PZ}{\mathrm Z}
\newcommand{\MW}{M_\PW}
\newcommand{\MZ}{M_\PZ}
\newcommand{\Mt}{M_\mathrm{t}}
\newcommand{\MH}{M_\mathrm{H}}
\newcommand{\Mc}{m_\mathrm{c}}
\newcommand{\ccbar}{\ensuremath{c\bar{c}} }
\newcommand{\QQ}{\ensuremath{\mathrm{Q^2}} }
\newcommand{\Gmu}{ {G_\mu}}
\newcommand{\GF}{ {G_F}}
\newcommand{\eV}{\unskip\,\mathrm{eV}}
\newcommand{\eVc}{\unskip\,\mathrm{eV}^2/c^4}
\newcommand{\MeV}{\unskip\,\mathrm{MeV}}
\newcommand{\GeV}{\unskip\,\mathrm{GeV}}
\newcommand{\TeV}{\unskip\,\mathrm{TeV}}
\newcommand{\fb}{\unskip\,\mathrm{fb}}
\title{History of accelerator neutrino beams}
\author{Ubaldo Dore \inst{1}
\and Pier Loverre \inst{1,2}
\and Lucio Ludovici \inst{2}\fnmsep\thanks{\email{lucio.ludovici@infn.it}} 
 }
\institute{Dipartimento di Fisica, Universit\`a di Roma Sapienza, P.le A.Moro 2, 00185 Rome Italy 
\and INFN, Sezione di Roma Sapienza, P.le A.Moro 2, 00185 Rome Italy }
\abstract{
Neutrino beams obtained from proton accelerators were first operated in 1962. 
Since then, neutrino beams have been intensively used in particle physics and evolved in many different ways. 
We describe the characteristics of various neutrino beams, relating them to the historical development of the physics studies and discoveries. 
We also discuss some of the ideas still under consideration for future neutrino beams.
} 

%
\maketitle
\section{Introduction}
\label{sec:intro}

In 1956 the detection by Reines and Cowan \cite{Cowan1956} at a nuclear reactor proved that the "desperate remedy" proposed by Pauli \cite{Pauli} more than a quarter of a century before, was a real sub-atomic particle. The particle postulated by Pauli to save the principle of energy conservation in beta decays, was named by Fermi neutrino and turned out to be a central ingredient of his new theory of weak interactions \cite{Fermi}\footnote{A previous manuscript on the same subject sent to Nature by Fermi was rejected by the Editor because ''it contained speculations too remote from physical reality to be of interest to the reader''}.

Three years later in a famous paper Bruno Pontecorvo discussed the hypothesis that neutrinos emitted in the decay of charged pions were different from the ones produced in beta decays \cite{Pontecorvo1959}.
The paper considered specific experiments to test this hypothesis, proposing also to use neutrinos from decay in flight of pions produced by accelerated protons hitting a target.

Independently, in 1960 Mel Schwartz published the first realistic scheme of a neutrino beam for the study of the weak interaction \cite{Schwartz1960}. 
His calculations showed that pions and kaons forward produced by high energy protons striking a target, would decay in to a collimated beam of neutrinos, intense enough to make their interactions detectable.

In 1962 the AGS accelerator at Brookhaven was commissioned and an experiment with the first neutrino beam was carried on. 
The experiment was designed to test the existence of two different neutrinos. 
It found that the neutrino interactions produced only muons. 
The absence of interactions with the production of electrons brought to the conclusion that $\Pnumu$ and $\Pnue$ were different particles \cite{Danby1962}.
In 1988, the Nobel Prize in Physics was awarded jointly to Leon M. Lederman, Melvin Schwartz and Jack Steinberger {\it ''for the neutrino beam method and the demonstration of the doublet structure of the leptons through the discovery of the muon neutrino''}.

In his Nobel Prize lecture, Schwartz acknowledged Pontecorvo's 
contribution to the question of two neutrinos \cite{Schwartz1988}:
{\it ''[...] we became aware that Bruno Pontecorvo had also come up with many of the same ideas as we had. He had written up a proposed experiment with neutrinos from stopped pions, but he had also discussed the possibilities of using energetic pions at a conference in the Soviet Union. His overall contribution to the field of neutrino physics was certainly major''}. 

After this first discovery many important experiments have been made, using neutrino beams with different characteristics. 

In this paper we describe the characteristics of various neutrino beams, relating them to the historical development of the physics studies and discoveries and to the existing literature.
The subject is divided in four parts, reflecting in an approximate way the evolution of the research in neutrino physics.

In section \ref{sec:begin} we describe the first beams, at Brookhaven and then at CERN, which defined many general aspects, later shared by a large majority of neutrino beams.

We dedicate section \ref{sec:1970-2000} to discuss the period 1970-2000 in which experiments at neutrino beams gave fundamental measurements to establish the electroweak theory, to study the structure of the nucleon, its constituents, the quarks, and the strong force between them.

In section \ref{sec:beam-osci} we discuss the neutrino beams used for oscillation experiments. 
Though searches for oscillation were done with neutrino beams before, a dedicated effort started after the discovery of atmospheric neutrino oscillation in the late 1990's.

Section \ref{sec:future} is devoted to the main alternative ideas for future neutrino beams which have been discussed in the last 20 years.

\section{The beginning of neutrino beams at proton accelerators}
\label{sec:begin}
\subsection{The two-neutrino experiment}
\label{sec:twoneutrino}

In 1960, new synchrotron accelerators, the AGS at Brookhaven and the PS at CERN had just completed their commissioning. 
These new accelerators, operating at record proton energies up to $30 \GeV$, opened the exciting possibility of studying neutrino interactions at the GeV scale, only a few years after the detection by Reines and Cowan of neutrinos from a nuclear reactor.

\begin{figure}[ht]
\centering
\includegraphics[width=1.0\textwidth]{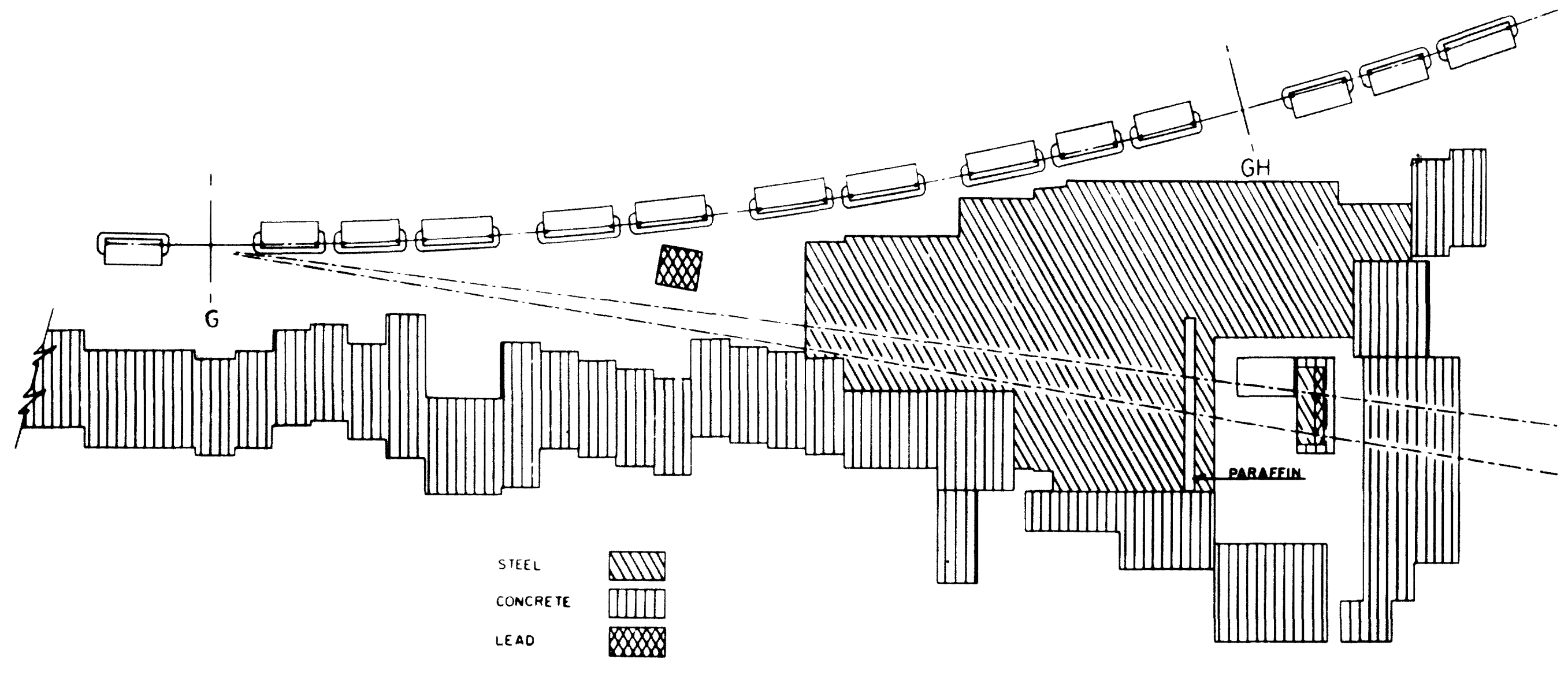}
\caption{Layout of the AGS 1962 two-neutrino experiment \cite{Danby1962}}
\label{fig:AGS}
\end{figure}

As mentioned in the introduction, the first experiment with accelerator neutrinos ran in 1962 at Brookhaven. The layout of the two-neutrino experiment is shown in figure~\ref{fig:AGS}. 

The $15 \GeV$ AGS proton beam struck a beryllium target placed at the end of a straight section of the accelerator, producing pions and a small fraction of kaons. 
Neutrinos were produced in the decay of pions and kaons drifting towards the detector (at a mean angle of $7.5^\circ$) in 21~m free space. 
All particles other than neutrinos were absorbed in a 13.5~m iron wall, shielding the 10~t aluminum spark chambers detector.

In an exposure of $3.5 \times 10^{17} $ protons, 56 neutrino interactions were identified. 
Out of these, 29 were assigned to single muon neutrino interactions while 6 shower events, which could possibly contain an electron, were compatible with expected backgrounds.
If neutrinos produced together with a muon in the $\pi \rightarrow \mu \nu$ or $\PK \rightarrow \mu \nu$ decays, were the same as those produced in the beta decay $\Pn \rightarrow \Pp e \nu$, the experiment would have detected about 29 neutrino interactions with an electron shower, in clear contrast with the observation.
The experiment then proved in an elegant way the existence of two different neutrinos, the $\Pnumu$ and the $\Pnue$.

\subsection{First focused beams}
\label{sec:firstbeams}

The Brookhaven experiment was the first to brilliantly exploit the forward production of pions and kaons at a high energy proton accelerator, in order to get an intense flux of neutrinos, and to use it to prove the existence of two types of neutrinos.

Remarkably, the experiment also measured a number of events for the quasi-elastic reaction $\nu \Pn \rightarrow \mu \Pp$ (called elastic at the time) compatible within $30 \%$, with what expected from the estimated neutrino flux combined with the theoretical cross-section \cite{Neut63}.
In spite of neutrino cross sections of the order of $10^{-38} \mathrm{cm^2}$ at $1 \GeV$, the available proton intensities (up to $10^{12} \mathrm{protons/s}$) and the estimated yields of pions and kaons, made the detection of neutrino interactions possible.

The limited statistics collected by the Brookhaven experiment (of the order of 100 events in a 10 tons detector for 1 month data taking), also showed that to study neutrino interactions in more detail, improvements in the beam design, in particular to increase the intensity, would have played an important role. 

The idea of an experiment to solve the two-neutrino puzzle was considered also at CERN in competition with Brookhaven. 
However the experiment designed at CERN appeared unable to collect a sufficient statistic and it was cancelled in favour of constructing a new beam.
The construction of the neutrino beam at CERN started in 1961 and the beam became operational in 1963 \cite{Ramm1963A,Ramm1963B,Giesch1963}, just one year after the Brookhaven experiment. 
With respect to the beam at Brookhaven, there were two important advances: the extraction of the proton beam and the focusing of the secondary mesons.
An efficient extraction system \cite{Kuiper1959,Kuiper1965} allowed to send the proton beam to an external target, instead of having the target positioned inside a straight section of the accelerator. 
With an external target, the trajectories of the charged mesons produced in the target are not disturbed by the fringe field of the accelerator magnets, and there is space to surround the target with magnets, in order to focus the charged secondaries toward the experiment. 
The focusing of pions and kaons is fundamental for the intensity of the neutrino beam, even if the angular distribution of the neutrinos in the decay produces an unavoidable dispersion also in a completely parallel beam of mesons. 

For the purpose of focusing, Simon van der Meer proposed in 1961 the new idea of the ''horn'' magnet.
He writes: \textit{"Divergent beam of charged particles can be made nearly parallel by a magnetic horn that is analogous to an internally reflecting conical surface in geometrical optics"} \cite{Meer1961}.

\begin{figure}[ht]
\centering
\includegraphics[width=1.0\textwidth]{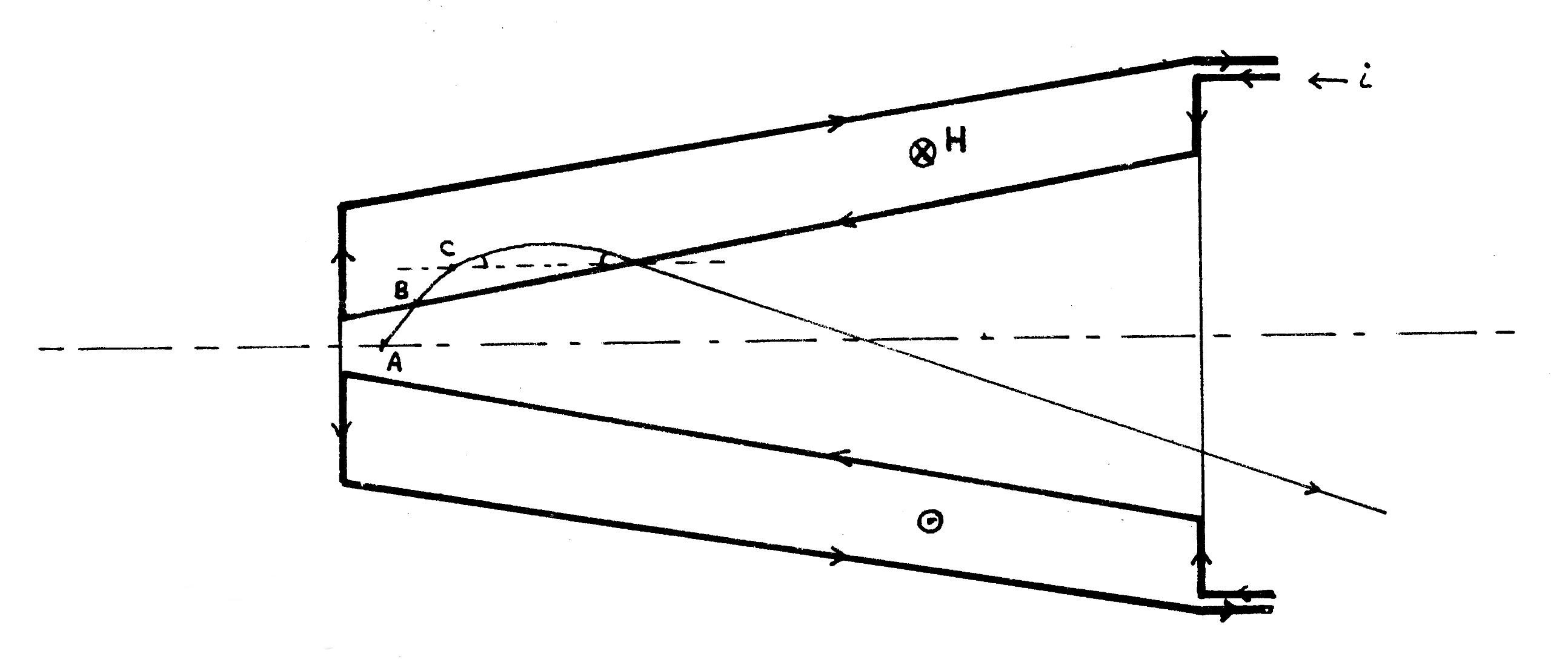}
\caption{ Van der Meer's original sketch of a magnetic horn \cite{Meer1961}}.
\label{fig:horn1}
\end{figure}

Figure~\ref{fig:horn1} shows the drawing of the van der Meer's horn.
It consists of two metallic conductors shaped as truncated cones, disposed around the axis of the proton beam. 
The inner and outer surfaces are connected at the basis and a current flows through the internal surface and returns through the outer surface, forming a current sheet of rotational symmetry. 
The resulting toroidal magnetic field is confined in the volume between the inner and outer conductors and the field lines are circles around the axis.
The field intensity goes like 1/r with r the distance from the axis. 
Particles entering the volume between the inner and outer conductor receive a kick from the magnetic field which focus them toward the horn axis for a broad momentum range.
Since the toroidal field focus particles of one sign and defocus those of the opposite sign, by choosing the direction of the current it is possible to get a neutrino beam (from $\pi^+$) or an antineutrino beam (from $\pi^-$).

\begin{figure}[ht]
\centering
\subfloat{\includegraphics[width=.5\linewidth]{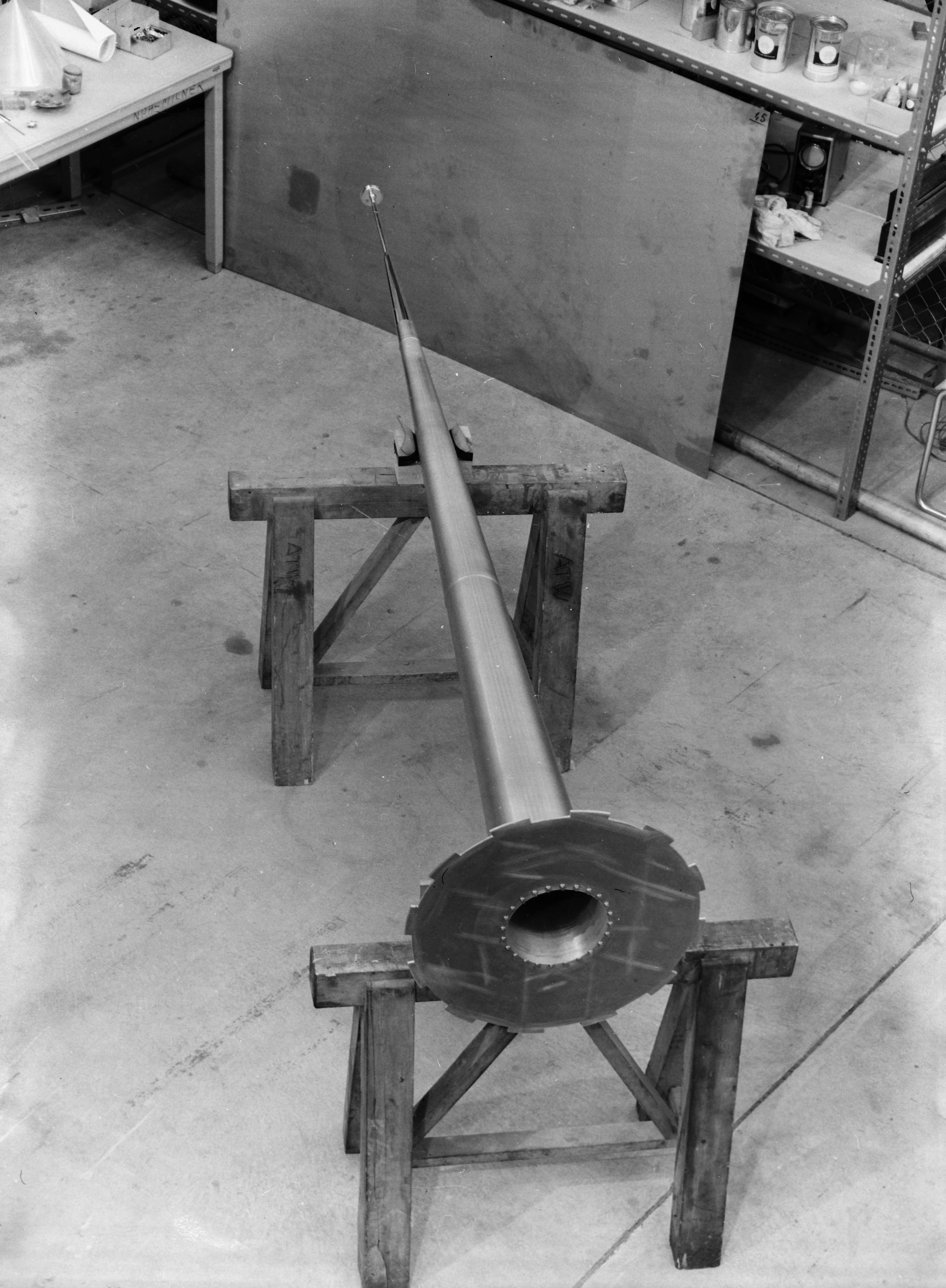}}
\subfloat{\includegraphics[width=.494\linewidth]{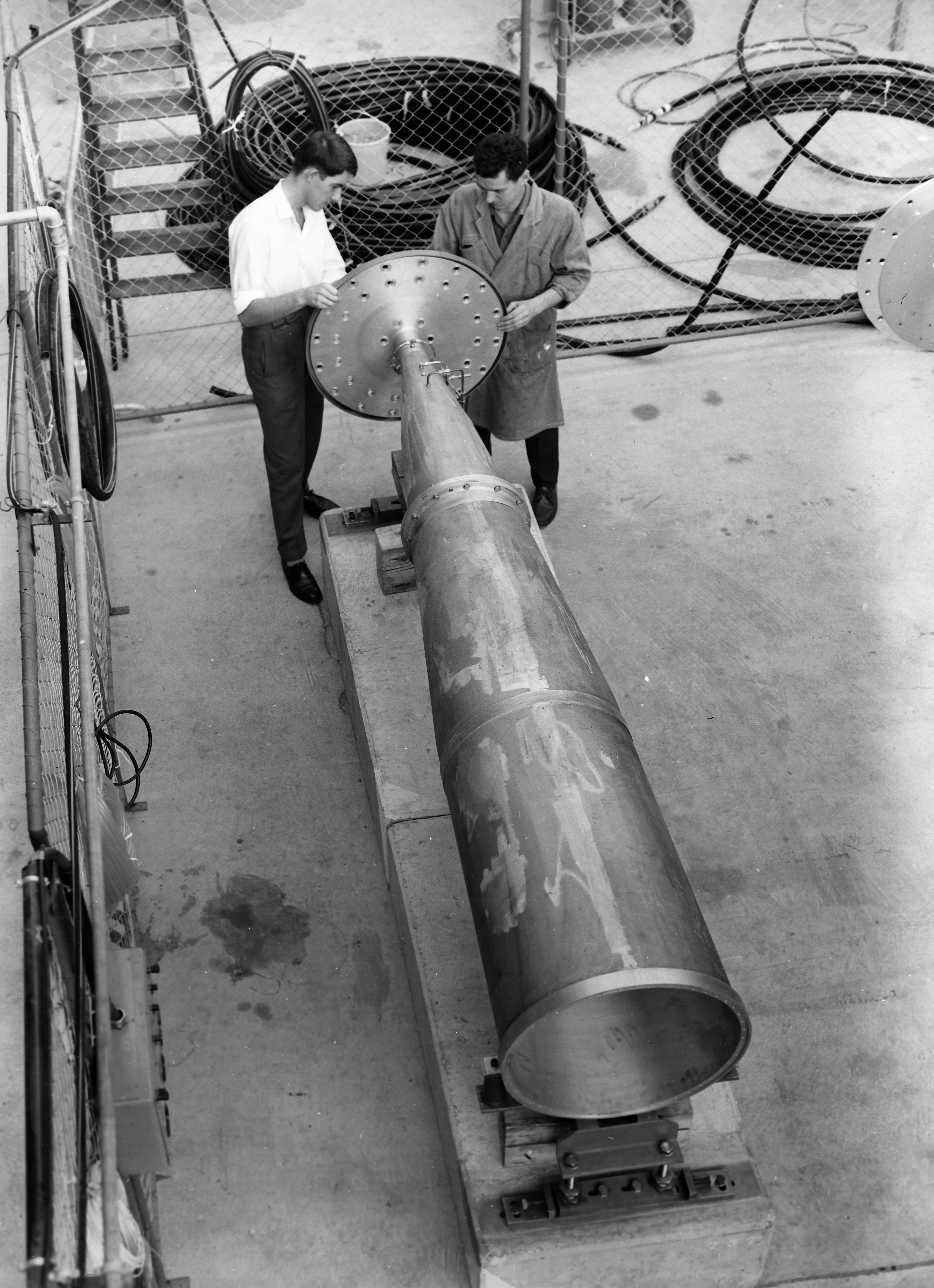}}
\caption{The magnetic horn constructed at CERN in 1963: (left) inner conductor, (right) outer conductor (Courtesy of CERN).}
\label{fig:hornCern}
\end{figure}

The first magnetic horn was constructed for the CERN neutrino beam at the $25 \GeV$ PS, and started operation in 1963 \cite{Ramm1963A,Ramm1963B,Giesch1963}.
A picture is displayed in figure~\ref{fig:hornCern}, which also illustrates the origin of the name ''horn''. 
Because of the high current (300 kA) and the need to dissipate the heat, it was necessary to operate the horn in pulsed mode. That was made possible by the fast extraction, designed to send the proton beam on the external target within a short pulse of $2.1 \mathrm{\mu s}$, the PS revolution time, repeated every 3 s.
The construction has to be very robust in order to withstand the magnetic forces repeated over an order of $10^7$ pulses. At the same time the conducting surfaces have to be thin, to reduce interactions and energy losses of the charged particles traversing them.

The beam operation of the horn was very successful, with an estimated gain of a factor 20 in intensity with respect to the bare target configuration. 

Van der Meeer's horn has been the first focusing system adopted in neutrino beams. Soon after, a similar horn, but with a different system for generating the current in the inner conductor, was constructed for the neutrino beam at the $12.5 \GeV$ ZGS (Zero Gradient Synchrotron) at Argonne National Lab \cite{Vogel65}.

In the years, the basic idea of a magnetic lens made by a toroidal field between two conducting surfaces of axial symmetry has been adopted in the design of many neutrino beams around the world, including all experiments presently running or under construction. 
The most relevant evolution of the horn concerns the shape of the conductors and the use of multiple horns in cascade.
We will mention these aspects later. 
Here, we note that a three horns system was already adopted in the 1967 beam at CERN. 
In that beam, the horn was followed few meters downstream by two more toroidal magnetic lenses, called {\it reflectors}. 
Particles well focused by the horn passed without deflection through the central hole of the reflectors, while over- or under-focused particle entered the magnetic field of the reflector, which corrected their trajectories. 

\subsection{Evaluation of the neutrino flux and energy spectrum}
\label{sec:flux}

\begin{figure}[ht]
\centering
\includegraphics[width=0.8\textwidth]{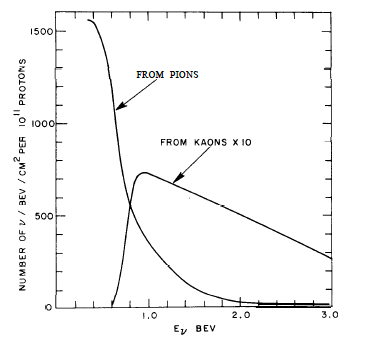}
\caption{Energy spectrum of neutrinos expected for the AGS running at $15 \GeV$ \cite{Schwartz1988}}.
\label{fig:BNLbeam}
\end{figure}

The first experiments at BNL and at CERN based their estimate of the neutrino flux on the available measurements of pions and kaons production, complemented by phenomenological models of hadroproduction. 
Such calculations are clearly affected by large uncertainties, but allow nevertheless realistic estimates. 
We recalled already that the two-neutrino experiment at BNL found a rate of neutrino interactions in agreement within $30 \%$ with the Fermi theory prediction. 
This is also mentioned by Mel Schwartz in his Nobel lecture \cite{Schwartz1988}, from where we reproduce the calculated energy spectrum of neutrinos shown in figure \ref{fig:BNLbeam}.

A new method to determine the flux of neutrinos as a function of the protons on target (POT), was first implemented at BNL in 1965 \cite{Burns65}. 
The method, which became largely common afterwards, consists in putting counters at different depths and different radial position, in the shielding which follows the decay region.
By taking into account the energy loss of muons in the filter, it is possible to connect the decrease of the muon flux as a function of the depth, to the muon energy spectrum.

The neutrino energy spectrum is related to the measured spectrum of the accompanying muons in the decay but there are two important limitations. 
The flux of low energy muons ($E < 1 \GeV$) cannot be measured because they are absorbed in the first section of the dump, where they are embedded in the hadronic showers produced by the undecayed pions and kaons. 
To reconstruct the neutrino flux one has to know the fraction $\PK/\Ppi$ as a function of the energy, since the parent particle of the muons is not known from the measurement.

The technique of measuring the muon flux in the dump, introduced in 1965 at Brookhaven, was adopted at CERN in 1967, for the horn focused beam.
Continuous monitoring was performed using mainly ionisation chambers. The analysis, taking into account new hadroproduction measurements and detailed tracking through the triple horn focusing system, led to determine the neutrino flux with a precision of $15\%$ for $1.5 < E < 2 \GeV$, of $10\%$ for $2 < E < 5 \GeV$, and of $20\%$ for $E > 5 \GeV$ \cite{Wachs69}. 

We note that hadroproduction data are  always needed to calculate composition, spectrum and intensity of neutrino beams.
We will discuss at the end of section \ref{sec:wideband} the earlier hadroproduction experiments done with single-arm spectrometers and at the end of section \ref{sec:earlyosci} the more recent hadroproduction experiments with full acceptance detectors.

\subsection{First years of operation}
\label{sec:firstres}

The neutrino beams developed in the 1960's at Brookhaven, CERN, and ANL did not immediately produce results of the same importance as the initial BNL experiment. While the detection of neutrino interactions was a success by itself, statistics of few hundreds events did not yet allow deep insight in the structure of the nucleon or detailed tests of the weak interaction.
Perhaps the hottest topic for the early experiments has been the search for the W boson, the supposed mediator of the weak interaction
\cite{Lee1960A,Lee1960B}.
Essentially no prediction of the value of the W mass was then available. Lee only quotes $\mathrm{m_W \geq m_K}$ because of the absence of the $\mathrm{K\rightarrow W + \gamma}$ decay. 
However, the energy of the neutrino beams at BNL, CERN, and ANL allowed to explore only a small interval of the possible $\mathrm{m_W}$ value, above $\mathrm{m_K}$.

As an example, let us consider the data collected at CERN in a 0.75 t heavy liquid bubble chamber and in a multi-ton spark chamber detector, exposed in 1963 and 1964 to the horn focused beam. 
Besides confirming with larger statistics and better details the results of the Brookhaven experiment ($\Pnue$ interactions were also detected, with rates in agreement with what expected from kaons decay), the experiment quote lower limits of the order of $1.5 \GeV$ on the W boson mass \cite{Block1964,Bienlein1964,Bernardini1964B}.

One may also note, that at a conference in 1964, Bernardini introduced his report on the CERN neutrino experiment by saying \textit{''If something was really good in the experiment, it was the beam''} \cite{Bernardini1964A}.
The remark, besides celebrating the innovations of the CERN neutrino beam, may reflect a certain disappointment for the results of the experiment. 
In more recent times, looking back at that period, some authors discussed the hints for the linear rise of the neutrino-nucleon cross-section, and for the existence of the neutral weak current, which could have been already observed in the early experiments, \cite{Veltman2003}, \cite{Perkins2013}.

However, as we will see in section \ref{sec:1970-2000}, it is by the late 1960's that changes in the physics scenario, the beginning of operation of higher energy accelerators and the construction of much bigger detectors, opened the way to 20 years of precision measurements and discoveries in the field of weak interactions and of the nucleon structure.

\subsection{Neutrinos from stopping pions}
\label{ssec:stopping-pions}

Before coming to discuss in the following paragraphs the developments of high energy neutrino beams, we shall mention that experiments have also used stopped pions, an idea also originally discussed by Pontecorvo, to get neutrinos with a well known spectrum in the few tens of MeV range.
Neutrinos from the decay of pions at rest (DAR) are obtained when protons hit a target large enough that pions come at rest before decaying.
Then, negative pions are absorbed, while positive pions decay with the sequence $\pi^+\rightarrow \mu^+\nu_\mu$ followed by $\mu^+\rightarrow e^+\APnumu\Pnue$. 
The flavour composition and the spectrum of neutrinos are remarkably different from the decay in flight.
Neutrinos are produced isotropically and their spectrum has a prompt monochromatic component of 30~MeV $\nu_\mu$ and delayed $\APnumu$ and $\Pnue$ components, corresponding to the Michel spectra of the muon decay at rest.
Experiments at stopped pion neutrino sources were LSND at Los Alamos (LANSCE) and Karmen at ISIS (RAL), both powered by 800~MeV proton accelerators of respectively 0.1~mA, 0.2~mA intensity. 
They produced measurements of cross section for neutrino interactions on carbon \cite{LSND,Karmen} and they are well known for one of the most controversial result on neutrino oscillation, known as ''LSND anomaly'' \cite{LSND-anomaly}.
Neutrino cross sections on different nuclei in this energy range are relevant for understanding core-collapse dynamics and nucleosynthesis in supernovae and this motivated further proposals of neutrino from stopped pions at neutron spallation sources. 
Instantaneous neutrino fluxes as high as $1.7 \times 10^{11} \nu_\mu /\mathrm{cm}^2/\mathrm{s}$ obtained at the SNS facility in Oak Ridge have allowed the recent observation of the coherent neutrino-nucleus scattering \cite{coherent}.

\section{Neutrino beams for weak interactions and neutrino-nucleon inclusive scattering (1970-1990)}
\label{sec:1970-2000}
\subsection{The context at the end of the 1960's}
\label{ssec:context60}

The end of the 1960's represents a turning point for the physics with neutrino beams. In a few years between the end of the 1960's and the beginning of the 1970's, various facts gave the start to more than 20 years of very important investigations in the field of weak interactions and of the structure of the nucleon.
 
In 1968 the electron-proton scattering experiment at SLAC \cite{ep1,ep2} discovered that at high momentum transfer, the form factors were independent of $\QQ$. 
This \textit{scaling} behaviour characteristic of scattering off point particles \cite{bjorken}, was soon interpreted in terms of a composite structure of the protons.
The constituents are the quarks, the half unity spin, fractional charge particles originally proposed independently by Gell-Mann and Zweig \cite{gellmann,zweig} as a tool to provide an ordering scheme for hadrons. 
With the recognition in 1973 of the asymptotic freedom by D. Gross, D. Politzer and F. Wilczek (2004 Nobel Prize winners) \cite{gross,politzer}, QCD emerged as the elegant new gauge theory of strong interactions, able to make reliable perturbative calculations in the high energy regime. Deep inelastic lepton scattering, both with charged leptons as well as with neutrinos, provided a crucial means of testing the prediction of the new theory and gave to it the first experimental support.

In the same years the separate works of S.L. Glashow \cite{glashow}, A. Salam \cite{salam}, and S. Weinberg \cite{weinberg} completed the construction of the electroweak theory, which unifies the weak and electromagnetic interactions in a common framework, consistently applicable up to the highest energies. 
Then in 1971 t'Hooft, proved that the theory is renormalizable \cite{thooft}. 

These revolutionary changes in particle physics, coincided with the opening in the early 1970's of a new energy domain for the neutrino beams. In 1972 started to operate the $350\mbox{--}400 \GeV$ proton accelerator at Fermilab. 
In 1972 was commissioned the $70 \GeV$ proton accelerator at Serpukhov, and in 1976, the $300 \GeV$ Super Proton Synchrotron (SPS) at CERN. 

Neutrino beams were among the top priorities for all major proton accelerators, since high energy neutrinos appeared to be the ideal probe both in the field of the weak interactions and for investigating the structure of the nucleon.

On one side, the newly defined picture of the electroweak interaction made a firm prediction, the existence of the neutral current, fundamental for the correctness of the theory. 
In high energy neutrino interactions, the prediction implied the existence, besides the charged current reaction $\Pnumu N \rightarrow \mu+\mathrm{hadrons}$, also of the neutral current reaction $\Pnumu N \rightarrow \Pnumu+\mathrm{hadrons}$.
The search for the neutral current and its discovery in neutrino interactions in 1973, was the first of a series of extremely important results on the electroweak interaction obtained with neutrino beams.

At the same time, high energy neutrinos immediately appeared specially well suited to continue the exploration of the structure of the nucleon initiated by the SLAC e-p experiment. This in particular because, contrary to charged leptons, neutrinos scatter differently on quarks and antiquarks, owing to the V-A character of the weak currents.

The new field of research also lead to the development of new massive electronic detectors. 
The role of big electronic detectors is better understood by considering the basic reaction studied to investigate the structure of the nucleon. 
This is the deep inelastic charged current scattering of $\Pnumu$ and $\APnumu$ on the nucleon: $\numupbarp N \rightarrow \mu^{\tiny\pM}\mathrm{hadrons}$. 

The scattering can be treated in an inclusive way, i.e. regardless to the configuration of the hadron system resulting from the fragmentation of the scattered quark. 
By measuring the momentum of the muon and the total energy of the hadrons $E_\mathrm{had}$, and considering that the direction of the incoming neutrino is known, the four-momentum conservation defines the full kinematics of the reaction. 
The kinematical variables usually adopted are $x$ and $y$, with $ x = \QQ / 2 m_p E_\mathrm{had} $ (with $x$ interpreted as the fraction of the nucleon momentum carried by the struck quark) and $y = E_{had}/E_{\nu}$.
The nucleon structure functions, and then the composition of the nucleon in term of valence and sea quarks, are derived from the study and comparison of the double differential cross sections $ d^2 \sigma / dx dy$ for $\Pnumu$ and $\APnumu$.

Tracking calorimeters, able to measure the energy and possibly the direction of the hadronic showers, complemented by muon spectrometers for the momentum and charge of the muon, are powerful instruments for these studies. 
In the early 1970's began the construction of such detectors with masses up of 1000 t, able to collect millions of events in the high energy beams. 

In parallel to the big electronic detectors, all along the 1970's and later, fundamental contributions also came from the very large bubble chambers, with their superior resolution of the details of the interactions. 
At CERN, while Gargamelle, a protagonist of the first hour, broke in 1979, BEBC, the $15 \mathrm{m^3}$ Big European Bubble Chamber, continued to operate until 1984. 
At Fermilab the 15-foot Bubble Chamber took data from 1973 to 1988. 
SKAT, the $7 \mathrm{m^3}$ heavy liquid bubble chamber constructed at Serpukhov, was used between 1974 and 1992. 

In the following sections, we will first describe the main features of the neutrino beams which came in use in the 1970's, and then relate the beams to the developments in the understanding of the electroweak interaction and of the quark structure of the nucleon.

\subsection{Wide band neutrino beams}
\label{sec:wideband}

The most common type of neutrino beam, in use at the proton accelerators of all laboratories, is the so-called Wide Band Beam (WBB). 
The basic principle consists in focusing pions and kaons over a wide interval of momenta, in order to maximise the overall neutrino flux.
The focusing of a WBB is usually obtained with different systems of toroidal magnetic lenses of the kind of the horn described in section \ref{sec:firstbeams}. An alternative to toroidal lenses (see e.g. \cite{Carey1971}) consists in using doublets or triplets of quadrupole magnets focusing in orthogonal planes. 
Compared to horns, this technique is less efficient, allows for a smaller acceptance, and requires additional magnets if sign selection is wanted.
On the other hand, design, construction and operation of quadrupole magnets is less challenging and less expensive, and the system is operated in continuous mode, contrary to the horn which need to be pulsed during the extraction of the proton beam. Moreover, the charged mesons do not have to cross metallic surfaces for bending. 
At Fermilab, various experiments have used quadrupole-focused beams, and quadrupole focusing has been used in the 1990's for the latest high energy neutrino beam at the $800 \GeV$ Tevatron, the SSQT (Sign Selected Quadrupole Train) beam, used by the NuTeV experiment\cite{Yu1998}. The SSQT beam will be described in the next section.

Besides quadrupoles and toroidal lenses, other focusing systems have been proposed and discussed. There are several interesting ideas but they have found little application \cite{Kopp2007}.

Going back to the focusing with toroidal lenses, we recall that the construction of the first horn has been followed by extensive studies aimed at optimising the shape of the conductors for the need of the experiments. The study of the shape of the surfaces is essential, since the shape defines the field strength encountered by a particle as a function of its production angle and production point. 
In the design, optical analogies are helpful, but must then take into account the spread in the transverse momentum distributions of pions and kaons, that a fraction of the particles is produced in secondary interactions, and that the source of particles is not point-like , but it is given by the length of the target (usually of the order of 1m).
Here we also note that the WBB is not the only type of beam to employ horn focusing. In fact, horn systems of the appropriate configuration and design, allow to obtain neutrino beams with very different energy spectra.
This is particularly relevant for neutrino oscillation experiments, as will be mentioned in the following sections.

\begin{figure}[ht]
\centering
\includegraphics[width=1.0\textwidth]{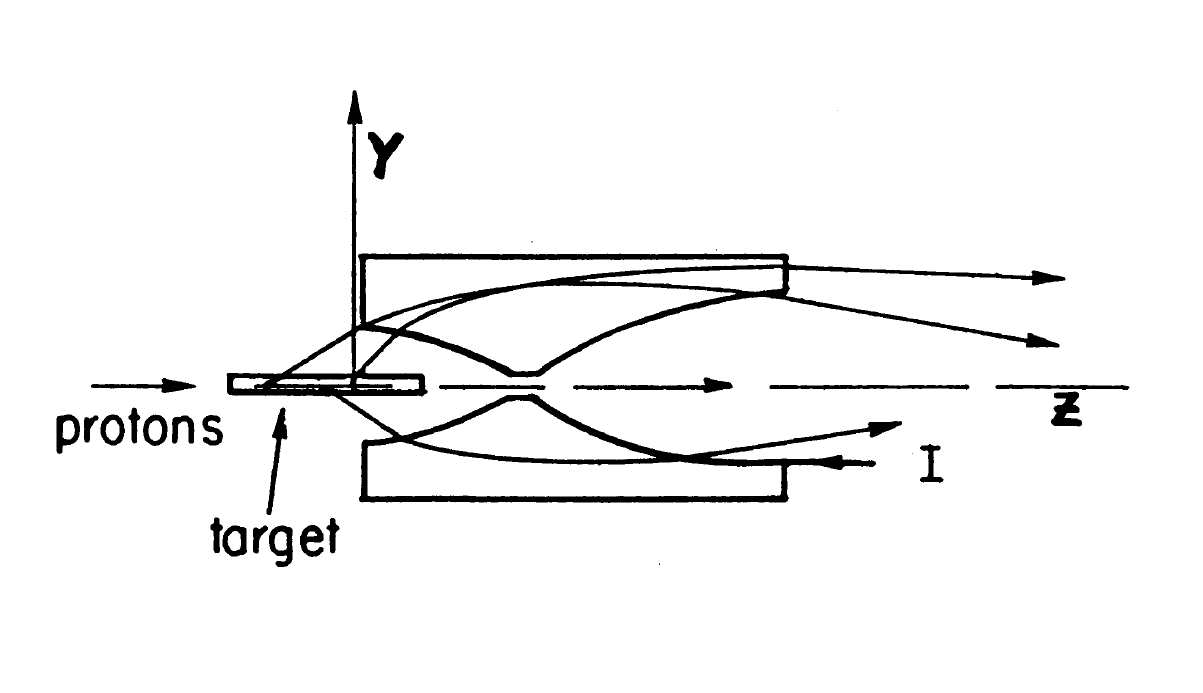}
\caption{ Sketch of a parabolic horn \cite{Klein1978}}
\label{fig:horn2}
\end{figure}

Concerning the toroidal lenses for a WBB, a design often adopted, different from the conical horn of Van der Meer, is the parabolic horn sketched in figure~\ref{fig:horn2}.
We also recall that horn focused WB beams, make often use, downstream to the first horn, of one or two additional toroidal lenses, with larger central holes, to correct trajectories of over- or under-focused particles, as already mentioned in section \ref{sec:firstbeams} for the CERN 1967 beam. 

We shall now look in more detail the characteristics of a horn focused WB neutrino beam.

\begin{figure}[ht]
\centering
\includegraphics[width=\textwidth]{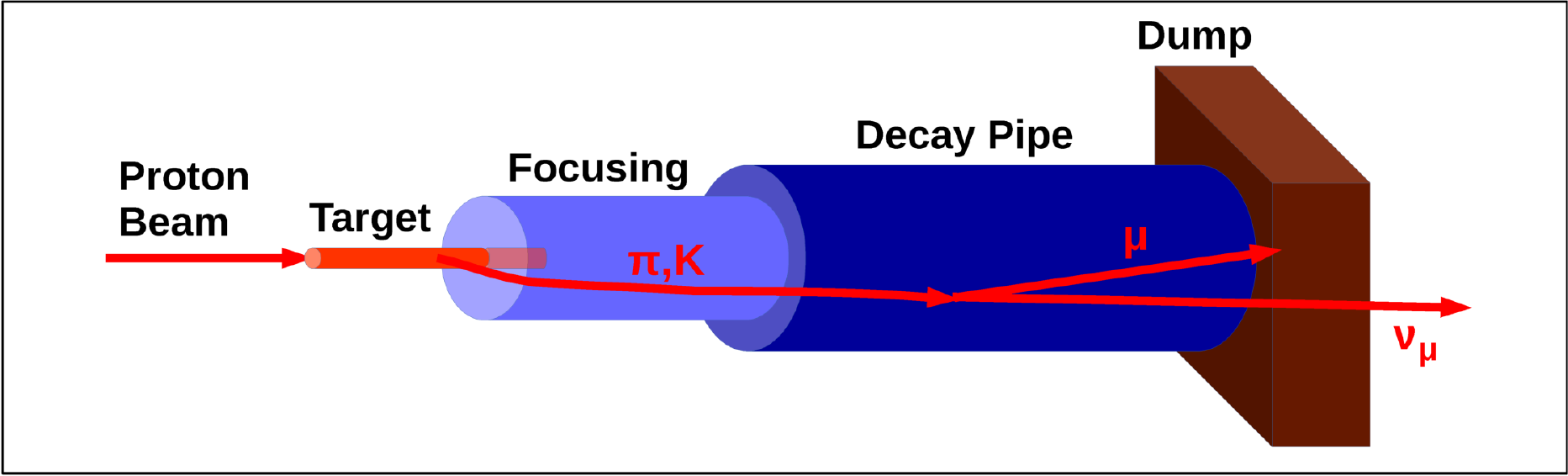}
\caption{ Sketch of a typical neutrino beam }
\label{fig:nubeam}
\end{figure}

An elementary sketch of the beam line is in figure~\ref{fig:nubeam}.
Protons are extracted from the accelerator and steered to hit an external target to produce pions and kaons. 
Usually, targets are rods few interaction lengths long (order of one meter) and few millimetres in diameter. The target is followed by (or inserted in) focusing magnets. 
The beam of charged pions and kaons enters then the decay region, an evacuated or helium filled tunnel where the long lived mesons can decay.
The length of the tunnel vary for the different experiments and has an influence on the energy spectrum and composition of the beam, owing to the different decay length of the charged pion and kaon (e.g., at 10 GeV, $L_{decay} (\pi^\pm) = 560m$ and $L_{decay} (K^\pm) = 75m$).
Finally, at the end of the decay tunnel a dump of iron and/or rock is used to absorb hadrons and electrons, and to range out the muons.

Note that in figure~\ref{fig:nubeam} the various elements are not in scale. 
For example, the decay tunnel for the CERN-Gran~Sasso beam of years 2000 is almost 1~km long \cite{cngs}, while the decay region of the two-neutrino experiment in 1962 was 21~m long. Approximately 300~m of rock are needed to range out a $200 \GeV$ muon.

Horn focused WBB are mainly composed by muon neutrinos, coming from the dominant $\mu \nu_\mu$ decay mode of the $\pi^\pm$ (b.r. $100 \%$) and of the $K^\pm$ (b.r. $64 \%$). 
They are $\Pnumu$ or $\APnumu$ beams, depending on the charge of the parent hadrons selected by the horns. Muon neutrinos coming from pions and kaons of charge opposite to the chosen one, are present in small fraction. They are called ''wrong-sign'' neutrinos. Small fractions of $\Pnue$'s and $\APnue$'s are also present.

\begin{figure}[ht]
\centering
\includegraphics[width=0.9\textwidth]{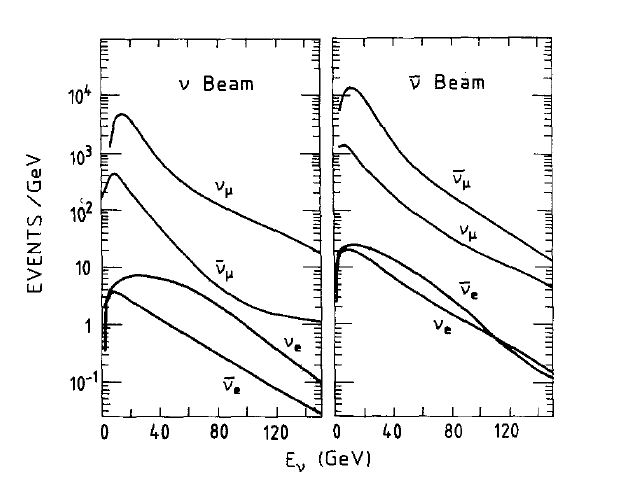}
\caption{Energy spectra for the neutrino species in the CERN WBB with the SPS 400 GeV proton beam. Left: neutrino beam. Right: antineutrino beam. The overall normalisation is arbitrary and different for the two figures \cite{Doren1989}}
\label{En-WBB}
\end{figure}

An example of the energy spectrum of the different components of a typical WBB is given in figure~\ref{En-WBB}. The figure refers to the horn focused WBB obtained with the $400 \GeV$ protons of the CERN SpS, and used by the BEBC, CDHS and CHARM experiments in the 1980's.
The figures show a neutrino beam (left) and an antineutrino beam (right). 
In the figures, the overall normalisation is arbitrary and different for the two.
Because pions and kaons are produced from pp or pn collision, i.e. the initial state has positive charge, positive mesons are more abundant than the negative ones.
It follows that the neutrino beam is more intense than the antineutrino beam, and that the contamination of wrong-sign neutrinos is higher (up to $10 \%$) in the antineutrino beam.

The contribution of $\Pnue$'s and $\APnue$'s is small, usually of the order of one percent each. It comes mainly from three sources. One source is the secondary decay of the muons produced in the decay of pions and kaons.
Then there is the contribution of $\mathrm{K_{e3}}$ decay of charged kaons (b.r. $4.8\%$). The last relatively important contribution is due to the $\mathrm{K_{e3}}$ decay of neutral kaons.

The evaluation of intensity, energy spectrum and composition of a WBB is primarily based on the monitoring of the flux of muons with counters positioned at different depths and different radial positions, in the dump which follows the decay tunnel.
As described in section \ref{sec:flux}, this technique has been introduced at Brookhaven in 1965, and since then adopted in all laboratories.
The characteristics of the neutrino beam are inferred from the measurements of the muon flux, by combining them with models for the production of pions and kaons in the target. 

The models for the production of pions and kaons in the target are based on external information from ancillary experiments, compilations of hadroproduction measurements at different proton energies and on different targets. 
The  hadroproduction experiments are usually done with single-arm spectrometers, taking data on a limited set of angles and momenta and use targets different in material and shape from those used in neutrino beams.
Phenomenological models are then used to extend the measurements to the full range of angles and momenta of the secondaries, and to extrapolate the hadroproduction data taken on different materials to the material of the target of a specific neutrino beam.

Several experiments, using small acceptance single-arm spectrometers taking data at different angles and momenta, were done to measure the production of secondaries on various target materials.  
These measurements cover the range of proton energies from the $10\mbox{--}20 \GeV$ of the ZGS at Argonne \cite{cho}, the PS at CERN \cite{allaby} and the AGS at Brookhaven \cite{abbott}, up to $300\mbox{--}450 \GeV$ of the proton synchrotrons at Fermilab \cite{baker} and CERN \cite{atherton,spy}. 

With the growing interest in neutrino oscillation in recent years, dedicated experiments have been designed for full acceptance measurements of pion and kaon production. They will be discussed at the end of section \ref{sec:earlyosci}.

We conclude this section by recalling that WBB are specifically designed for maximising intensity.
Other neutrino beams, described in the next section, allow a more precise evaluation of the neutrino flux and composition. 
These beams, in general less intense than the WBB, are therefore useful for those studies (e.g. the measurement of total cross sections) where statistical errors are less relevant than a precise knowledge of the beam.

\subsection{Dichromatic, Narrow Band and SSQT (Sign Selected Quadrupole Train) beams}
\label{nbb}

An alternative to the WBB described in the previous section was first developed in 1974 at Fermilab, with the construction of the so-called dichromatic beam \cite{Limon1974}. In that dichromatic beam, the 300 GeV protons struck a cylindrical target not aligned with the decay tunnel. Following the target, a transfer line with dipole magnets and momentum slits selected pions and kaons in a given momentum bite. The mesons of the selected momentum were focused by quadrupoles, and guided to the decay tunnel. 
Neutrinos coming from the decay of pions and kaons of the same momentum show a dichromatic energy distribution, i.e. two separated broad distributions, corresponding to the two energy distributions of neutrinos from the $\Pnumu \mu$ decay of pions and kaons. 
Furthemore, in the two-body $\Pnumu \mu$ decay, at fixed momentum of the parent particle, the decay angle defines the energy of the emitted neutrino. The decay angle can be approximately inferred by the impact point of the neutrino in the detector, which therefore yields a measurement of the neutrino energy, up to the twofold ambiguity of the parent particle.

\begin{figure}[ht]
\centering
\includegraphics[width=1.0\textwidth]{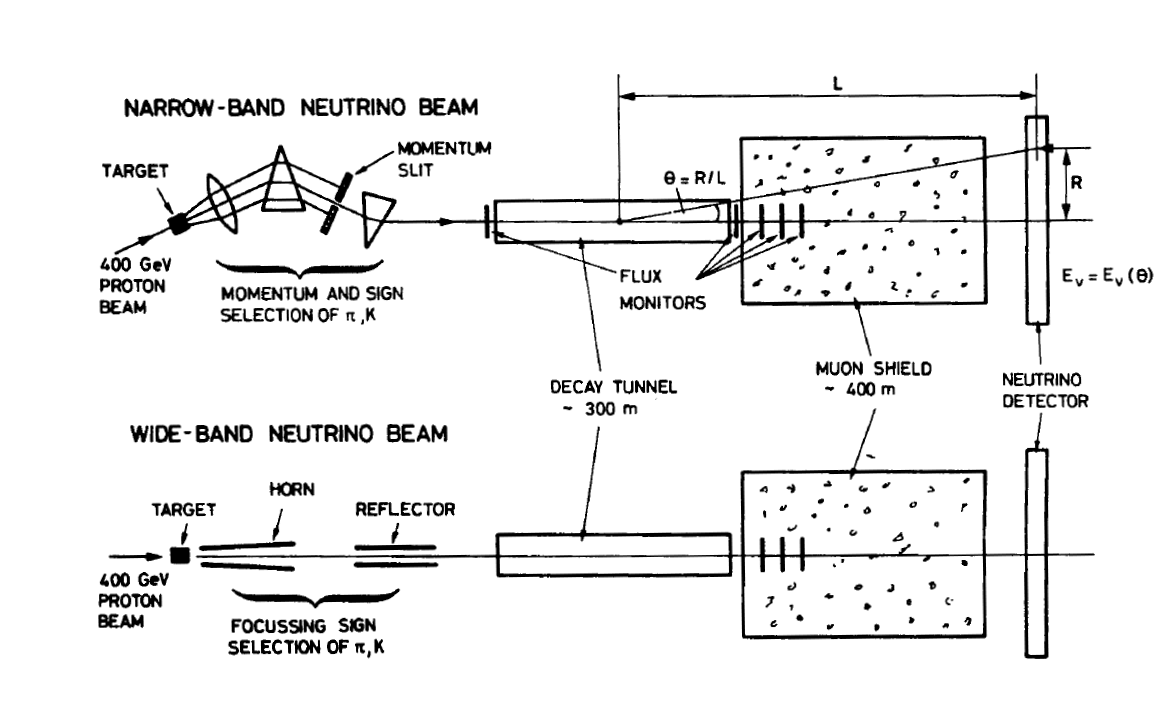}
\caption{ Layout of the CERN SPS WB and NB beams \cite{Steinberger1988}}
\label{fig:wbnblayout}
\end{figure}

\begin{figure}[ht]
\centering
\subfloat{\includegraphics[width=0.5\textwidth]{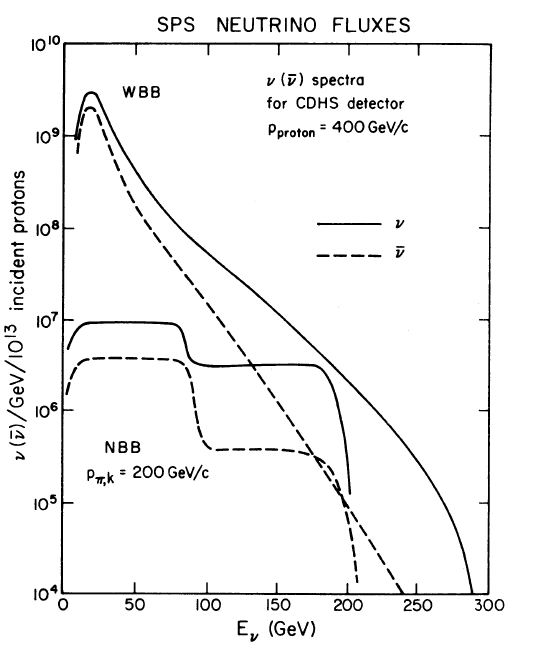}}
\subfloat{\includegraphics[width=0.465\textwidth]{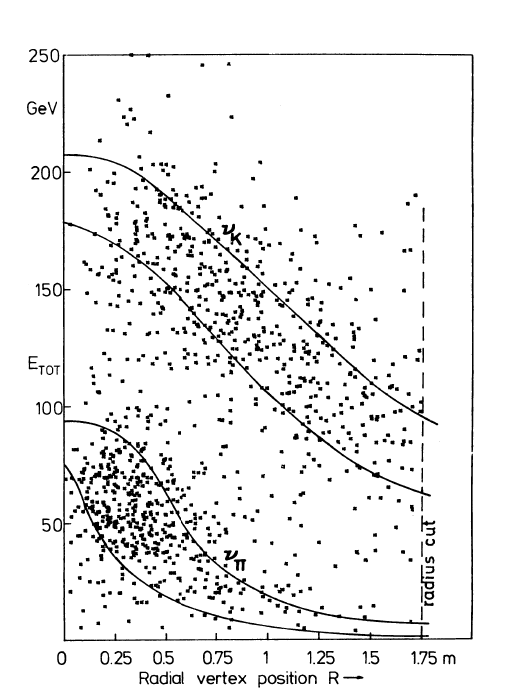}}
\caption{ \textit{Left}: Energy spectra of the CERN SPS WB and NB beams \cite{Steinberger1988}. \textit{Right}: E vs R for the CERN NBB \cite{Klein1978} }
\label{fig:wbbnbb}
\end{figure}

In 1977 CERN started to operate a similar dichromatic beam, named Narrow Band Beam (NBB), because of the selection of the parent mesons in a narrow energy interval.
A sketch comparing the layout of the NB and WB beams at the $400 \GeV$ SPS is shown in figure \ref{fig:wbnblayout}, taken from
Steinberger's Nobel lecture\cite{Steinberger1988}.
From the same reference, we show in Figure~\ref{fig:wbbnbb}(left) the comparison of the energy spectra of the $\Pnumu$ and $\APnumu$ NB and WB beams. 
For the NBB, pions and kaons were selected in a $9 \%$ momentum bite around 200~GeV.
The NBB spectra show the double structure, corresponding to the decay of pions and kaons. 
It is seen that at the peak, the intensity of the WBB is more than two order of magnitude greater than that of the NBB, while the intensities become comparable at the high energy edge.
Figure~\ref{fig:wbbnbb}(right) \cite{Klein1978} illustrates the previously discussed relation between the neutrino energy and the radial position of the event, for $\Pnumu$ charged current interactions in the CDHS detector exposed at the CERN SPS NBB. 
In the plot, the energy of the neutrino is obtained by adding the measured energies of the muon and of the hadronic shower.
The plot shows that the energy of the interacting neutrino can be inferred, within the twofold ambiguity, from the radial position of the interaction vertex in the detector. 
This is an important piece of information for neutral current events, where the energy of the incoming neutrino cannot be derived otherwise, since only the energy of the shower can be measured. 

In spite of the lower intensity, many features make the NBB attractive with respect to the WBB.
Because of the bending of the mesons beam, the NBB has a much smaller contamination of wrong-sign neutrinos (neutrinos from the decay of mesons of the wrong charge), and of neutrinos from neutral kaons. 
The bending of the beam and the selection in momentum allow higher precision in the monitoring of the flux of the parent mesons, and 
also, in some configuration, to directly measure the $\Ppi/\PK$ ratio, e.g. with differential Cherenkov counters\cite{Berge1987}. 
Then, the known kinematics and branching ratios of the decaying mesons make it possible to predict with good accuracy the intensity, energy spectrum and composition of the resulting neutrino beam.
\cite{Yu1998,nutev}.
\begin{figure}[ht]
\centering
\includegraphics[width=1.0\textwidth]{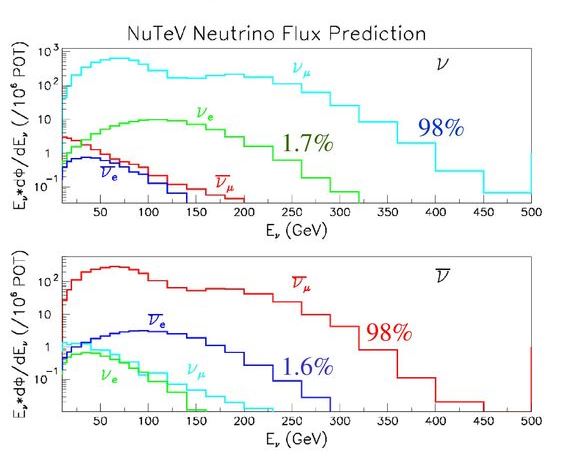}
\caption{Energy spectra multiplied by E$_\nu$ for the four neutrino species of the $800 \GeV$ SSQT neutrino beam \cite{naples}}
\label{fig:ssqt}
\end{figure}

Another important type of beam, combining characteristics of the WB and NB beams, is the Sign Selected Quadrupole Train (SSQT) which was in operation at Fermilab in the second half of the 1990's for the NuTeV experiment. The SSQT beam is the last evolution of the neutrino beams operated at the $800 \GeV$ Tevatron since 1984. Mesons produced by the $800 \GeV$ protons were sign selected by a dipole magnet with very large momentum acceptance ($250 \pm 100 \GeV$) and then focused by three successive quadrupoles \cite{naples}. 
Figure \ref{fig:ssqt} shows the energy spectra for the different components of the beams (note that the flux is multiplied by the neutrino energy). The spectra extend to the highest energy, without the steep energy fall-off, typical of the WBB, and the wrong-sign contamination is much smaller than in a WBB.

\subsection{Neutrinos from a beam dump}
\label{dump}
The usual neutrino beams are formed by the decay in flight of pions and kaons produced in the interaction of a high energy proton beam in a long and thin target. 
An alternative consists in dumping the proton beam in a dense block extending also in the transverse direction. In this way, most of the high energy pions and kaons will interact before decaying, and will not produce high energy neutrinos.
Short lived particles instead, like for instance charmed mesons with proper lifetimes of the 
order of $10^{-13}$~s, will decay before interacting, originating a flux of high energy neutrinos.
For example, in copper, a $20 \GeV$ pion has a decay length of $11 \times 10^4$cm, to be compared with an interaction length of $\sim 15 $cm. A 20~GeV $\mathrm{D^+}$ meson instead, has a decay length of 0.34~cm, i.e. 50 times smaller than the interaction length.
By absorbing pions and kaons in the dump, it is then possible to get a beam of neutrinos coming only from short lived particles, the so-called prompt neutrinos.

It is in fact soon after the discovery of the J/$\Psi$ in 1974 that beam dump experiments were proposed, aiming at the search of the production of short lived particles in proton-nucleon interactions, by looking at a signal of prompt neutrinos. 
The detection of the prompt signal is not straightforward. The non-prompt background of neutrinos from pions and kaons, although strongly reduced by energy cuts (e.g. analysing only neutrino interactions above $20 \GeV$), it is not entirely eliminated. Conventional neutrinos are suppressed by factors of the order of 1000, but the cross section for the production of charmed mesons by $400 \GeV$ protons is also smaller by similar factors. 
The basic technique developed to extract the prompt neutrinos signal consists in repeating the measurements with targets of different density. 
This is done by using segmented targets, e.g. solid slabs alternated with air gaps, or targets of different materials. 
While the flux of prompt neutrinos is essentially independent from the density of the target, the background from pions and kaons decays decreases with increasing density.
The prompt signal can then be obtained by extrapolating the measurements at different densities, to {\it infinite} density, where the non-prompt background is zero.

The first beam dump experiments were made at CERN in 1977 and 1979 in the first years of operation of the $400 \GeV$ SPS, and at the $70 \GeV$ accelerator at Serpukhov in 1977 \cite{Asra78,Abram1982,Fritze80,Jonker80}. 
At CERN, after the first detection of a signal of prompt neutrinos with the 1977 run, the experiment was repeated in 1979 with better conditions. 
Data were taken by the BEBC, CHARM and CDHS detectors which were able to detect interactions attributed to prompt neutrinos, and deduce relevant information on charm production in proton-nucleon interactions. 

The CERN experiments also observed hints for a unexpected effect.
While the rate of prompt-neutrino interactions was compatible with the production of $c {\bar{c}}$ pairs, all three experiments observed a number of electron neutrino interactions smaller than the number of muon neutrino interactions. 
This was intriguing since in proton-nucleon interactions, the production of a $c {\bar{c}}$ pair and the subsequent decays should result in the four $\Pnumu$, $\APnumu$, $\Pnue$, $\APnue$ neutrinos all contributing with similar amounts to the total flux.
This because the mass difference between the light electron and muon leptons is negligible in the decay of the heavy c quark into the light s or d quark. 
Because of the large errors, the observed $\mathrm{e} / \mu$ ratio of $\sim 0.6$, violating universality, constituted only a 2-3 sigma effect. Nevertheless, it added a stimulus to perform a new series of beam dump experiments to improve the information about charm production and to possibly discover new sources of prompt neutrinos. 
However, the next generation of experiments in the early 1980's at CERN, Serpukhov and Fermilab found the ratio $\mathrm{e} / \mu $ to be in agreement with universality and with the characteristics of charm production better known by that time \cite{Berge1992,Doren1988,Duffy1988,Blum1992}). 

Another opportunity offered by beam dump experiments is the possibility to detect $\Pnutau$ interactions.
In the usual neutrino beams $\Pnutau$ and $\APnutau$ are practically absent, but they are expected to be produced in the decay of $D_s$ mesons, and therefore to constitute a sizeable fraction of the beam dump neutrinos. 
The beam dump experiment DONuT carried on in 2000 at Fermilab, used a nuclear emulsion detector and the $800 \GeV$ protons from the Tevatron for the first direct observation of tau neutrino interactions \cite{Kodama2001}. 
The main difficulty of the experiment resides in the necessity to maximise the acceptance for neutrino, by putting the detector as close as possible to the dump, therefore in a dirty environment, because of the high flux of muon. 
In DONuT the emulsion detector was at 36 m from the dump, and it was possible to detect 8 candidates of $\Pnutau$ interactions, with an expected background of 0.5 event.

The possibility of studying $\Pnutau$ physics at the LHC (Large Hadron Collider) was also investigated for sometime. 
Neutrino fluxes for beam dump experiments were considered, and it was also suggested to directly use the collimated flux of neutrinos coming from the decay of the charm and beauty particles abundantly produced at intersection regions \cite{DeRujula84,DeRujula93}. 
However, these ideas were not followed by any detailed project. 

Finally, we mention a beam dump experiment proposed at CERN in recent years. 
SHiP, \textit{A facility to Search for Hidden Particles at the CERN SPS} \cite{ship}, is an advanced project, possibly running at the 400 GeV SPS by 2025. 
Main purpose of SHiP is the search for \textit{hidden particles} supposedly created in the decay of heavy mesons produced in the dump. 
Detailed studies of $\Pnutau$ physics will also be possible. 

\subsection{Physics with neutrino beams in the 1970's and 1980's}

The new high energy beams described in the previous sections, began to operate in the early 1970's at Fermilab, Serpukhov, and CERN, with the new research programme on the structure of the nucleon and on the properties of the electroweak interaction, triggered by the electron-proton scattering experiment at SLAC, and by the success of the electroweak theory.

We shall however recall, that first results in these new domains were already obtained with the neutrino beam of lower energy at the CERN $28 \GeV$ PS, developed in the 1960's and described in section \ref{sec:firstbeams}. In fact, in 1970 started to take data the $12 \mathrm{m^3}$ heavy liquid bubble chamber Gargamelle. 
Data were collected both in the neutrino and antineutrino PS beam. 
Following the discovery by the e-p experiment at SLAC, the simultaneous study of the $\Pnumu N$ and $\APnumu N$ cross sections gave strong evidence for the existence of quarks inside the nucleon, and early support to QCD, with its property of asymptotic freedom \cite{GargaDIS}.

Neutral current events \cite{GargaNC1}, i.e. purely hadronic events attributed to the NC reaction $\Pnumu N \rightarrow \mu+hadrons$, were also observed for the first time in Gargamelle, with the PS beam.
The main challenge of this study was to distinguish purely hadronic final states induced by neutrinos, from interactions of neutrons, since neutrons produced in neutrino interactions in the upstream material, entered the detector together with the neutrino beam.
A detailed recollection of the discovery of the neutral currents, which included the observation in Gargamelle of a single electron event from the neutral current elastic scattering on electrons $\Pnumu e \rightarrow \Pnumu e$ \cite{GargaNC2}, is given by in \cite{Haidt}.

Immediately after Gargamelle announcement, the neutral currents were confirmed by other neutrino experiments in the US, in particular with neutrino beams at the 350~GeV accelerator at Fermilab \cite{fnalNC1,fnalNC2,fnalNC3}, where studies of the nucleon structure with more than ten times higher energies, were just beginning. 

Then, already in 1975, Fermilab announced the detection in neutrino interactions of di-muon events\cite{Benvenuti1975}, first evidence for the production of charmed particles. At the same time the first open charm, a charmed baryon, was observed in a hydrogen bubble chamber by the group of Samios, with the BNL neutrino beam \cite{samios}).

In the following 15 years, neutrino experiments have lead the studies of the nucleon structure, its composition in terms of quarks and gluons, finding also evidence of QCD effects, like scaling violations. 
In parallel, they made fundamental contributions to the electroweak theory, with the determination of the electroweak couplings, and in particular with the measurement of the weak mixing angle \sstw.

In those years, main protagonists were the  experiments using the big electronic detectors, with masses of hundreds of tons, and taking data with the neutrino beams from the 300-500 GeV accelerators at Fermilab and CERN,  and from the 800 GeV Tevatron, which started in 1984. 
We recall the CCFRR, FMMF and CCFR collaborations at Fermilab, and the CDHS and CHARM at CERN. 

A detailed review of the experiments and physics results of that period goes beyond our scope. 
We refer to the reviews in \cite{Diemoz,Conrad1998,Dore} and to the references therein, for the results obtained on structure functions and for the precision measurements in the electroweak sector. 
The book \textit{Neutrino Physics} edited by Winter in 1991 \cite{Winter}, collects monographs on different subjects, and reproduces also a few interesting historical papers. 
The Nobel lecture of Steinberger \cite{Steinberger1988}, though directed to a more general audience gives an excellent and rigorous summary of the physics with high energy beams in those years.

Going back to the beams described in the previous paragraphs, we note that horn focused WB beams were used at CERN, Brookhaven and Serpukhov. NB beams were intensively used at CERN and Fermilab, where also quadrupole focused WB beams operated. A few experiments used beam dump neutrinos, as discussed in section \ref{dump}.

Different beam configurations were often adopted to study the same subject.
However, the different characteristics of the WB and NB, allow for an approximate classification.
WB beams were employed for first order studies, and for the study of rare reactions. NB beams were mainly dedicated to precision measurements.

Examples of experiments with NB beams are the measurements of total cross-sections and the precise measurements of the electroweak mixing angle \sstw.

An important example of experiment making use of the high intensity of the WB beam, is the study of the reaction $\Pnumu e \rightarrow \Pnumu e$. 
The process is particularly interesting because of its purely electroweak nature. Neutrino scattering on electrons gives rise to a single electron in the final state, and is very difficult to detect, 
since the process has a cross-section about 2000 times smaller than that of neutrino-nucleon scattering. 
In spite of the tiny cross-section, the E734 experiment, running in the 1980's with a horn focused beam at the 28 GeV AGS at Brookhaven, collected hundred of events in a liquid scintillator calorimeter, and performed high precision measurements of electroweak parameters \cite{Ahrens1990}.

Finally, we mention a special use of NB beams. As explained in section \ref{nbb}, in a NB beams it is possible to infer the total energy of the event, and therefore the energy of the incoming neutrino, from the radial position of the interaction point in the detector. 
It is therefore possible to know the energy of the neutrino also in neutral current events, for which this information is missing, due to the escaping neutrino.
The CHARM and CCFR experiments, respectively at CERN and at FNAL, have exploited this feature and succeeded to measure the nucleon structure functions using neutral current events \cite{Jonker1983},\cite{ncCCFR}.

Toward the end of the 1980's, neutrino experiments diminished their importance in the field of quark physics and of electroweak interactions, with the start of proton-antiproton colliders at CERN in 1983 and at Fermilab in 1986, and of electron-positron colliders in 1989, first the SLC at SLAC, and soon after the LEP at CERN.  
Neutrino physics at the accelerators became oriented to the study of neutrino oscillations, making use, almost exclusively, of horn focused beam in various configurations. 
The study of neutrino oscillations with accelerator beams is described in the next chapter.

Note however that still in the mid of the 1990's, important results, mainly in the electroweak sector, were obtained by the NuTeV experiment, using the SSQT beam at the 800 GeV Tevatron of Fermilab \cite{naples}. 

\section{Neutrino beams for oscillation experiments}
\label{sec:beam-osci}
\subsection{Neutrino oscillation}
\label{sec:osci}
Neutrino flavour oscillation is one of the main topics in neutrino physics and for the last three decades it has been one the most active research fields in particle physics.
The idea that neutrino had small but non-zero masses was pioneered in the seminal papers by Pontecorvo \cite{Pontecorvo1957,Pontecorvo1958} and Maki, Nakagawa and Sakata \cite{mns}. It is remarkable that even before the discovery of the second neutrino, Pontecorvo challenged the idea that neutrino were massless and proposed neutrino-antineutrino oscillation. He wrote \textit{if the theory of the two-component neutrino is not valid (which is hardly probable at present) and if the conservation law for the neutrino charge does not hold, neutrino-antineutrino transitions in vacuum in principle would be possible} \cite{Pontecorvo1957}. 
According to the two-component neutrino theory, confirmed by the measurement of the neutrino helicity \cite{Goldhaber1958}, only the left-handed neutrino fields enter the weak interaction Lagrangian. 
This was considered an argument in favour of massless neutrino, though there are no general principles, like gauge-invariance for the photon, that require the neutrino to have zero mass.

If neutrino have masses, the flavor neutrino states are in general linear superposition of the neutrino mass states.
In the case of only two neutrinos, the two flavor states $\nu_\alpha$ and $\nu_\beta$ are linear combinations with a mixing angle $\theta$ of the two mass states $\nu_1$ and $\nu_2$. Since they have in general different masses $m_1$ and $m_2$, the two mass states evolve in time with different phases.
A neutrino is produced and detected only through weak interactions and so it is in a definite flavor state both when it is produced as well as when it is detected. If a neutrino is produced in a definite $\nu_\alpha$ flavor state, for instance $\Pnue$ in a beta decay or $\Pnumu$ in a pion decay, the probability to detect it in the different flavor state $\nu_\beta$ after it propagates for a distance $L$ it is given by the well known oscillation probability:
\begin{equation}\label{eq:probosci}
P(\nu_\alpha\rightarrow\nu_\beta) = \sin^2 (2\theta) \sin^2\left(1.27\frac{E[\GeV]} {L[\mathrm Km]}\Delta m^2[\eV^2]\right)
\end{equation}
where $E$ is the neutrino energy, $\Delta m^2=m_2^2-m_1^2$ is the mass squared difference of the two neutrino mass states and the numerical factor takes in to account the units used.

An oscillation experiment at a neutrino beam aims to detect variations with $L/E$ of the beam flavor composition. Short-baseline neutrino oscillation experiments, are usually defined by the ratio $L/E \lesssim 10$ $\mathrm{Km}/\GeV$ and are sensitive to $\Delta m^2 \gtrsim 0.1 \eV^2$. The observation of neutrino oscillation with small $\Delta m^2\lesssim 10^{-2} \eV^2$ requires long-baseline experiments, with a ratio $L/E \gtrsim 100$ $\mathrm{Km}/\GeV$.

The oscillation phenomenology is well explained today in the framework of the so-called Neutrino Standard Model ($\nu$SM). 
The three active neutrino flavour states, $\Pnue$, $\Pnumu$ and $\Pnutau$ are superpositions, through a unitary transformation, of the neutrino mass states, $\nu_1$, $\nu_2$ and $\nu_3$ with respective masses $m_1$, $m_2$, $m_3$. 
The mass matrix in the flavour basis is diagonalised by a $3x3$ unitary matrix, the so-called Pontecorvo-Maki-Nakagawa-Sakata (PMNS) matrix, usually written in terms of three mixing angles and a CP violating phase. 
Those are the parameters used to describe the neutrino oscillation phenomenology, together with the two independent mass-squared differences.

The current picture of neutrino oscillation is characterised by large mixing angles, strikingly different from the pattern of the small mixing angles of the CKM quark mixing matrix, and two small mass-squared differences. 
This picture emerged in the last three decades of the 20th century.
The convincing proof neutrino are massive and mixed particles was given by the Super-Kamiokande measurement of the up-down asymmetry in the atmospheric neutrino flux \cite{SK1998} and by the SNO measurement of the total neutrino flux from the sun \cite{SNO2002}, recognised jointly by the 2015 Nobel Prize to T. Kajita and A.B. McDonald.

The first experimental hint of neutrino oscillation appeared in the solar neutrino deficit measured by Davis Jr in the Homestake experiment \cite{Davis1968}, and, about twenty years later, in the early hint of the atmospheric neutrino deficit by the Kamiokande collaboration \cite{kamiokande-atmo}.
As the deficit of the solar neutrino flux was confirmed by the Kamiokande \cite{kamiokande-solar}, GALLEX \cite{GALLEX} and SAGE \cite{SAGE} experiments, with different techniques and energy detection thresholds there were little doubts on the experimental result. 
Still, until the SNO measurement in 2001, it was a hot debate in the scientific community whether this so called ''solar neutrino problem'' was a manifestation of neutrino oscillation or rather of the inadequacy of the solar model used to predict the neutrino flux.

On the other hand the existence of the atmospheric neutrino deficit was initially considered controversial, with inconsistent measurements by other groups, until the 1998 Super-Kamiokande confirmation. Also, if neutrino oscillation were the origin of the atmospheric deficit, they required a very large mixing angle, consistent with maximal mixing, $45^\circ$, while at that time small mixing angles, in analogy with the CKM mixing for quarks, were considered more natural \cite{Harari1989,Ellis1992} and neutrinos with masses in the range $1\mbox{--}100\eV$ were a popular dark matter candidate \cite{frampton}.

\subsection{Early oscillation experiments at accelerators}
\label{sec:earlyosci}

Before the design and construction of long-baseline neutrino beams, specifically optimised to search for oscillation with small neutrino mass squared differences, early oscillation searches were performed at the available neutrino beams and detectors. All these searches were short-baseline experiments, aiming to small mixing angles and mass differences of $0.1\mbox{--}100 \eV^2$, much larger than the ones we know today. They exploited, often with minimal, sometime with more relevant modifications, neutrino beams and detectors designed and optimised for other physics topics.
These experiments can be characterised as {\it disappearance} or {\it appearance} experiments. Disappearance searches look for the deficit of a neutrino flavor compared to the expected neutrino beam composition. Appearance searches look for a neutrino flavor not present in the beam or in excess with respect the beam composition.

Various experiments in the late 1970's and 1980's carried out searches with negative results and put limits on $\Pnumu$($\APnumu$) disappearance, on the disappearance of $\Pnue$($\APnue$), on the appearance of $\Pnue$($\APnue$) in $\Pnumu$($\APnumu$) beams, on the appearance of $\Pnutau$($\APnutau$) in $\Pnumu$($\APnumu$) beams. 
An old review of the status at that time and the developments of the experimental searches for oscillations at the major accelerator facilities can be found in \cite{chen} and \cite{bilenky}.

Searches for $\Pnue$ appearance are limited by the uncertainty in the intrinsic $\Pnue$ contamination in the beam. An experiment at CERN, PS191, \cite{ps191}, repeated by E816 at BNL \cite{e816}, reported a $\Pnue$ excess that was not later confirmed and was assumed to be affected by systematic errors not properly accounted in the simulation. 

In the late 1970's E531, a hybrid experiment using a target of nuclear emulsion, was built at Fermilab primarily to study charmed particles production in neutrino interactions. The experiment exploited the high spatial resolution of the nuclear emulsion also to search for short lived $\tau$ leptons produced in $\Pnutau$ charged current interactions. 
Since the contamination of $\Pnutau$ in a muon neutrino beam is practically negligible, this technique allows high sensitivity searches for $\Pnutau$ appearance, only limited by other backgrounds, mainly hadronic interactions wrongly reconstructed as $\tau$ decays \cite{Ushida1986}.

In the 1990's the CERN WB, horn focused neutrino beam at the SPS, was refurbished for two short-baseline experiments, CHORUS, using the same hybrid technique of E531, and NOMAD, searching for $\Pnumu\rightarrow\Pnutau$ oscillation with sensitivity to small mixing angles (at $\Delta m^2\sim 10\MeV$, a value much larger than the one we know today).
Despite the atmospheric neutrino deficit suggested a different scenario, a large fraction of the community was still unease with the idea of large mixing angles.
Quoting \cite{Ellis1992}: "\textit{The general see-saw model also suggests that neutrino mixing angles are related to quark mixing angles, which is also consistent with the oscillation interpretation of the solar neutrino data, and suggests that the forthcoming CHORUS and NOMAD experiments at CERN have a good chance of observing $\Pnumu-\Pnutau$ oscillations}".
Nature decided for a $\Delta m^2$ much lower and these experiments ended up with limits on neutrino oscillation \cite{chorus-final,nomad-final}

It is worth noticing that the short-baseline searches for neutrino oscillation with mass squared differences in the $0.1\mbox{--}10\eV^2$ range are still an active research topic today, with an experimental programme starting at the neutrino beam of the Fermilab booster \cite{fermilab-sbn}, using three detectors with baselines from 110 to 600 meters. This is motivated by experimental hints suggesting neutrino oscillation beyond the PMNS paradigm, in particular the existence of sterile neutrino mixed with the three active neutrino flavors.

Due to the flavour composition of the conventional neutrino beams, many searches were for disappearance of $\Pnumu$, looking for distortions in the measured muon neutrino spectrum with respect to the expectation in absence of oscillation, or appearance of $\Pnue$. The use of a second detector located upstream the neutrino beam, ''near'' the target, allows to compare the spectra measured by the two detectors and reduce the uncertainty from neutrino beam simulation, greatly improving the experimental sensitivity over a broad range of mass-squared differences. 

Dedicated oscillation experiments with two detectors were realised by E701/CCFR at Fermilab \cite{e701} and by CDHS \cite{Dydak1984} and CHARM \cite{Bergsma1984} at CERN. 

E701 consisted of a near and a far detector at distances of about 700 and 1100 meters, exposed to the narrow-band dichromatic neutrino beam at Fermilab, with neutrino with energy between $40$ and $230 \GeV$. To get a better control of the systematic, the events in the close and far detectors were compared as a function of neutrino energy, taking advantage of the dichromatic beam properties and the relation between the radial position of the neutrino interaction and the neutrino energy in the narrow band beam.

CDHS and CHARM were taking data at the CERN SPS neutrino beam and in order to improve the oscillation sensitivity at low mass squared difference, a lower energy neutrino beam was designed specifically for this oscillation experiment, using $19.2\GeV$ protons extracted from the PS accelerator. 
Two smaller near detectors were added, about 130 m from the target.
Near and far detectors were both oriented at $22^\circ$ to the new neutrino beam since the existing massive far detectors, located at about 850m from the target, could not be rotated from their alignment to the CERN SPS neutrino beam. 

The extrapolation of the flux from the near to the far detector was considered a critical point and despite the lower yields it was decided to run with bare target, without focusing. Quoting \cite{Dydak1984}: \textit{''No magnetic focusing was done behind the target. The divergence of the neutrino beam was therefore much larger than the solid angle of either detector, so that in the absence of oscillations the $\Pnumu$ flux scaled as approximately $L^{-2}$}''.

The use of two detectors is typical in today oscillation experiments at accelerators but does not dispense from a precision understanding of the neutrino beam.
The flux uncertainty cancellation in the near to far ratio is only partial and there is no cancellation for neutrino cross-sections measured at the near detector which are essential, together with the neutrino flux, to calculate the event rate at the far detector.
To control at the few percents level the systematic on the near to far extrapolation of the neutrino there is need of refined neutrino cross-sections models, of a complete simulation of the neutrino beam line (including the primary proton beam, the target, focusing system and decay tunnel), and of leveraging external ancillary measurements to model the secondary hadron yields from the target.

In the last twenty years the need to improve the sensitivity to the oscillation parameters, motivated new experiments designed as ''ex-situ'' ancillary measurements of the hadroproduction off the actual targets used in neutrino beams. They use large acceptance spectrometers to measure in one go the full phase space, or a large fraction of it, rather than a grid of single points in momentum and angle as in the single-arm spectrometer experiments.
Data are taken with thin targets to study the hadroproduction cross-sections, and with replicas of the actual target used in a specific neutrino experiment, in order to have an additional control on the particle production due to re-interaction of the secondaries, before escaping the target.

These modern experiments are usually based on TPCs (Time Projection Chambers) but already in the past large acceptance pion production data were taken with hydrogen filled bubble chambers exposed to proton beams (see for instance \cite{mueck}). 
The actual copper target of the neutrino beam at the CERN PS was placed inside a heavy liquid bubble chamber and exposed to protons to measure the secondary yields and calculate the neutrino and antineutrino spectra \cite{bctarget}.
In more recent years, the HARP experiment \cite{harp} has measured hadroproduction with several targets at the CERN PS, including replicas of the targets of the K2K experiment at KEK and the neutrino beam at the Fermilab Booster.
The MIPP experiment \cite{mipp} at Fermilab has taken pion production data used by neutrino experiments based at the NuMI beam.
More recently the NA61/SHINE experiment, successor of NA49, has measured the hadroproduction yields off the replica targets of the T2K \cite{na61-t2k} and NO$\nu$A experiments. 

More details on the subject of hadroproduction, on the experiments and on the phenomenological parametrizations of secondaries production, in connection with the design of a neutrino beam and of oscillation experiments, can be found in \cite{Kopp2007}. 

\subsection{Long-baseline: first generation beams}
\label{sec:lbl-first}

Long-baseline neutrino beams produced at accelerators were already considered in the 1970's, well before the Kamiokande measurement of the atmospheric neutrino deficit, to search for small neutrino masses through oscillation. Indeed as a basic quantum interferometer, oscillation experiments can be sensitive to very small neutrino mass squared differences, provided the baseline is long enough respect to the neutrino energy.
In this paper \cite{Mann1977} the idea of a long-baseline neutrino beam is discussed and remarkably the authors come even further by anticipating the search for CP violation in long-baseline neutrino experiments \textit{It is perhaps worth mentioning again that any actual neutrino oscillation phenomenon [...] might conceivably provide another means of observing a CP violation and thus allow a new attack on that fundamental problem}. 
It is also worth noticing that, at the end of the 1970's, in the design of the Gran Sasso laboratory in Italy, the underground experimental halls were aligned toward CERN, in anticipation of possible neutrino beams.

In order to match the $\Delta m^2$ mass squared difference observed in the oscillation of atmospheric muon neutrinos, even a neutrino beam of energy smaller than $1 \GeV$ needs a baseline of several hundreds of kilometres. Since the flux scales with $L^{-2}$ the beam power must follow in order to get a meaningful event rate. 
One feature of longer baselines is that the neutrino beam passes through the Earth's crust, with significant matter effects on the oscillation pattern for the longer baselines.

The discussion around long-baseline neutrino beams moved forward in the 1990's, with definite proposals submitted at KEK in Japan and at Fermilab and Brookhaven in the US.
By the end of the decade, with the Super-Kamiokande announcement of the up-down asymmetry in atmospheric neutrinos, the case for a long-baseline beam to confirm the oscillation at a controlled neutrino source became stronger than ever.

The first long-baseline beam was commissioned in June 1999 in Japan for the K2K (KEK to Kamioka) experiment. 
A wide band neutrino beam of $1.3 \GeV$ average energy was obtained from the decay of secondaries produced by $12 \GeV$ protons extracted from the KEK synchrotron on an aluminum target and focused by two magnetic horns. 
The beam was directed toward the 50 kt Super-Kamiokande water Cherenkov detector, located 250 km away, in the Kamioka mines.
A GPS based system provided the synchronisation over the long distance between the beam extraction and the far detector.
The layout included a suite of detectors at a near location 300m downstream the target, including a 1kt water Cherenkov, scaled-down version of the Super-Kamiokande detector. 
The beam was operated until November 2004 and collected $0.9\times 10^{20}$ protons on target in neutrino polarity. The experiment observed $112$ muon neutrino events at the far detector, with $158.1^{+9.2}_{-8.6}$ expected in absence of oscillation. 
This confirmed at a man made, controlled muon neutrino source the oscillation seen in atmospheric neutrinos, measuring a mass squared difference in agreement with Super-Kamiokande. \cite{k2k2003, k2k2006} 

\begin{figure}[ht]
\centering
\includegraphics[width=0.65\textwidth]{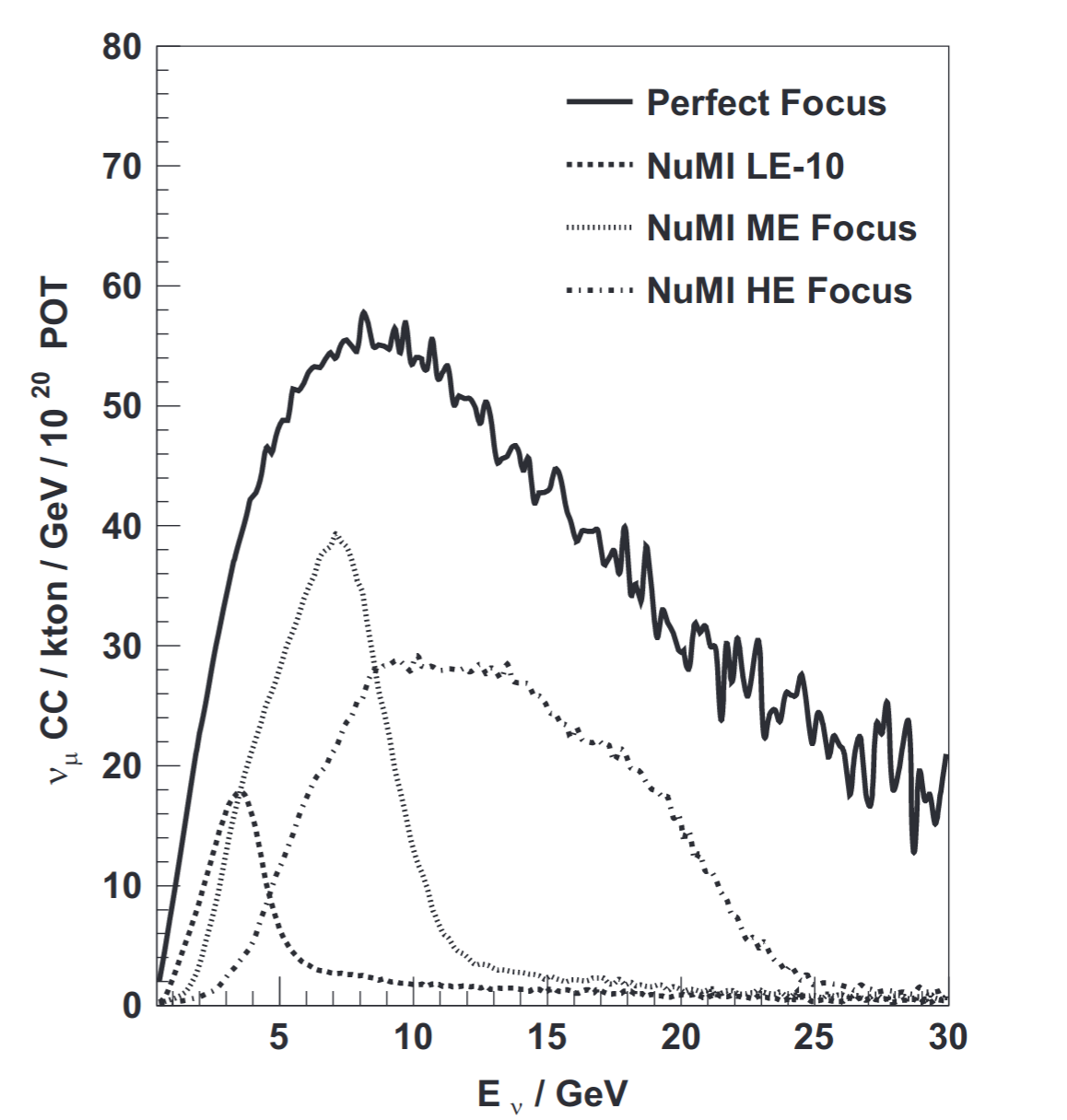}
\caption{Event rates (flux$\times$cross-section) for three different focusing schemes for the NuMI beam: low energy (LE), medium energy (ME) and high energy (HE), compared with the ideal perfect focusing \cite{numibeam}.} 
\label{fig:numi}
\end{figure}

In 2005, shortly after the end of K2K, a long-baseline beam, NuMI (Neutrinos from the Main Injector) was commissioned at Fermilab \cite{numibeam}. 
The beam was directed toward a newly constructed detector in the Soudan mines (Minnesota), MINOS (Main Injector Neutrino Oscillation Search), with a baseline of 735 km. 
At the time of the experiment design, mid-1990's, the range of possible $\Delta m^2$ values was quite broad, between $10^{-3}\mbox{--}10^{-1} \eV$.
Notice that the first measurement of the atmospheric neutrino deficit by Kamiokande \cite{kamiokande-atmo} pointed to a mass squared difference significantly larger than the value established later by Super-Kamiokande \cite{SK1998}.
A $120 \GeV$ proton accelerator like the Main Injector at Fermilab, can efficiently produce neutrino beams in the $1\mbox{--}10 \GeV$ energy range. 
With a baseline of several hundreds of kilometres this effectively cover the full $\Delta m^2$ range.
Once the baseline is fixed by the choice of the detector location, changing the neutrino beam energy is the only remaining flexibility to match $L/E$ for updated $\Delta m^2$ knowledge. 
The NuMI beam was designed with the flexibility to change the neutrino energy by using two magnetic horns with parabolic inner conductor profiles and adjusting the separation between them and the target to tune the momentum range of the secondaries focused. 
Three focusing schemes were optimised, with the neutrino spectra shown in figure \ref{fig:numi} in comparison with the perfect focusing, i.e. what could be ideally achieved by perfectly focusing all secondaries produced at the target.
The lowest energy configuration corresponds to the target inside the neck of the first horn and higher energies are obtained by retracting the target up to 2.5m from the horn. 
The target is mounted on a rail-drive system to fine-tune the beam energy.

By the time the beam line was completed and MINOS started taking data in 2005, the $\Delta m^2$ value measured in atmospheric neutrinos by Super-Kamiokande and then better determined by K2K, ended up to be at the lower bound of the range mentioned above and MINOS took data essentially only in the low energy configuration of the beam.
From 2006 to 2012 the experiment accumulated $10.7\times10^{20}$ protons on target in neutrino polarity, more than an order of magnitude larger than K2K, as well as $3.4\times10^{20}$ protons on target in anti-neutrino polarity.

In Europe the discussion of long-baseline projects started somewhat later, at the end of the 1990's, with the project of a new beam at CERN, directed toward the Gran Sasso laboratory in Italy, with a baseline of 732 km. 
While the main goal of the programmes in Japan and US was the muon neutrino disappearance, the European long-baseline experiment aimed to detect tau neutrino appearance in a muon neutrino beam.
The CNGS neutrino beam (CERN Neutrinos to Gran Sasso) used $400 \GeV$ protons extracted from the SPS accelerator to produce a neutrino beam focused by a two-horn magnetic system. 
To optimise for $\Pnutau$ detection, the neutrino beam had to be tuned to $17 \GeV$ average energy, well above the threshold for tau lepton production and an order of magnitude larger than, for instance, the NuMI beam which had a similar baseline. 
With respect to the K2K and the NuMI beams, which had $L/E$ matching the oscillation maximum for the atmospheric $\Delta m^2$, the rate of oscillated events at the CNGS is suppressed by almost two order of magnitude due to the term $\sin^2(1.27\Delta m^2 L/E)$ in the oscillation formula \ref{eq:probosci} with the expectation of a handful of events.
A major limitation of the CNGS with respect to the competing facilities is that there is no practical way to add a near detector.
The CNGS beam was commissioned in 2007 and operated until 2012 \cite{cngslast} and OPERA, a hybrid experiment using a nuclear emulsion target, reported the detection of its first $\Pnutau$ event in 2010 \cite{operafirst}.
Like others long-baseline experiments, OPERA used a GPS based system for timing and synchronization with the beam which allowed an experimental measurement of the neutrino velocity. 
In 2011 the announcement of faster then light neutrinos by the OPERA collaboration seized the attention of the media, until further cross-checks confirmed the result was due to an unfortunate experimental mistake. 
In the final analysis OPERA found 10 $\Pnutau$ events with $2.0\pm 0.4$ background events expected and \cite{operafinal}.

\subsection{Long-baseline: off-axis beams}
\label{sec:lbl-offaxis}

After the first generation of long-baseline experiments, a new programme started to chase the $\Pnumu\mbox{--}\Pnue$ sub-dominant oscillation and to study CP violation in neutrino oscillation and the ordering of neutrino mass.
The new long-baseline experiments, in Japan and US, are both using the so called off-axis technique to obtain an intense flux of neutrinos with a narrow energy spectrum, tuned at the desired value to maximise the oscillation probability.

In the off-axis technique, the secondary hadron beam is obtained as usual from proton impinging on a suitable target. Contrary to conventional beams, the target, the secondary beam optics and the decay tunnel are pointed a few degrees off the neutrino detector such that the detector itself is off-axis with respect to the neutrino beam. 
This idea was first proposed for a long-baseline experiment at Brookhaven \cite{Beavis1995} that was never realised.

The integrated neutrino flux is intuitively maximum when the detector is located on the beam axis where, for the dominant two body decay, the neutrino energy is directly proportional to the parent meson energy.
Since neutrinos are produced by the decay of mostly pions focussed into a nearly parallel beam by magnetic horns, the broad neutrino spectrum seen by a detector on axis is simply the reflection of the broad spectrum of the decaying pions.
 
\begin{figure}[ht]
\centering
\includegraphics[width=0.65\textwidth]{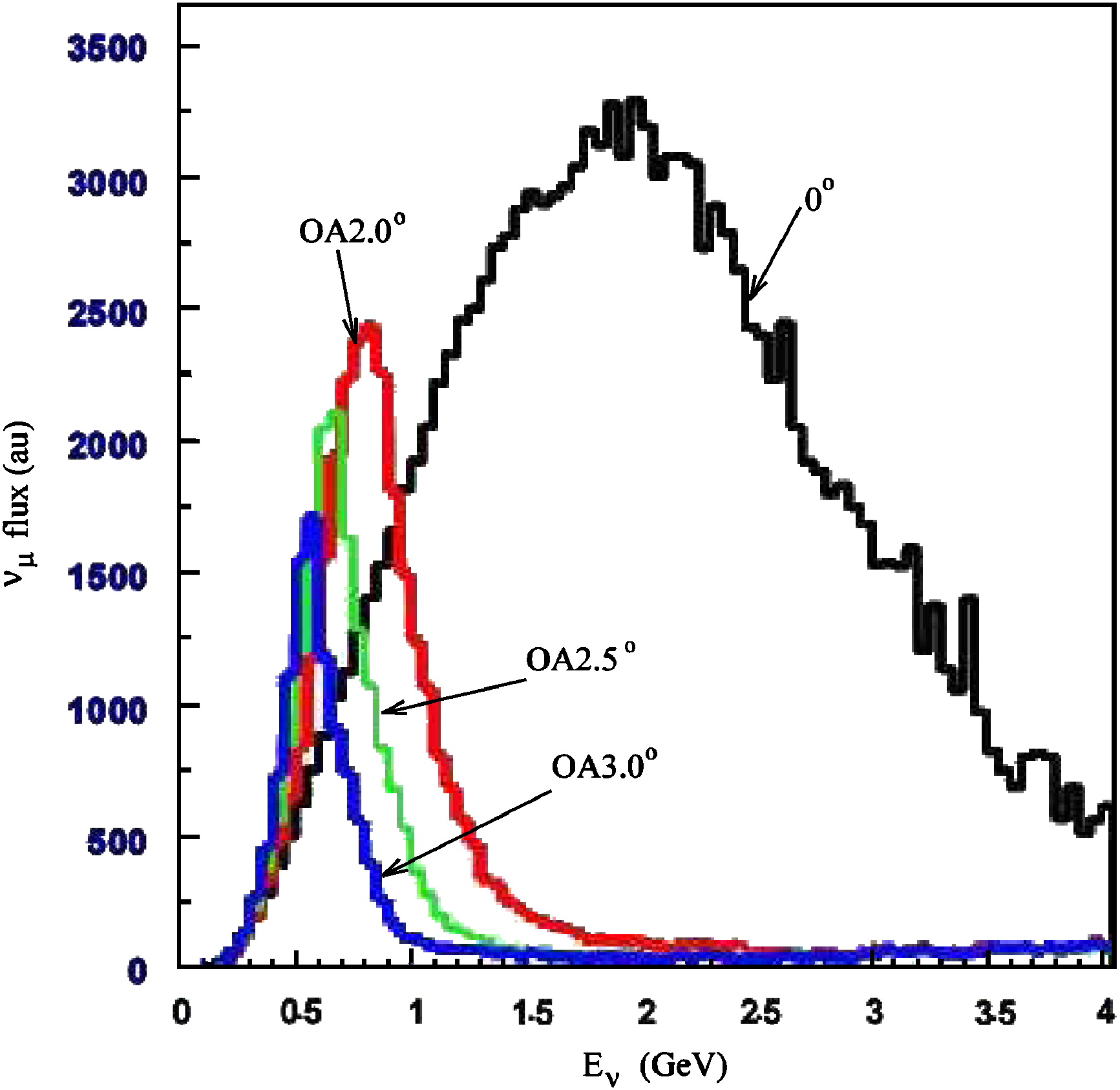}
\caption{Neutrino spectrum on axis and for different off-axis
angles for the T2K experiment \cite{Kudenko}}
\label{fig:t2koffspec}
\end{figure}

The neutrino energy and flux at a given angle $\theta$ with respect to the line of flight of pions of energy $E_\pi$ is derived easily from the pion decay kinematic.
For small angles ($\theta\ll 1$) and relativistic pions ($\gamma_\pi\gg 1$) one finds:
\begin{equation}\label{eq:offaxis}
E_{\nu}\simeq \frac{\mathrm{m}_\pi^2 - \mathrm{m}_\mu^2}{\mathrm{m}_\pi^2(1 + \gamma_\pi^2\theta^2)}\mathrm{E}_\pi \qquad \Phi_\nu \simeq \frac{1}{\pi \mathrm{L}^2}\left(\frac{\mathrm{E}_\pi}{\mathrm{m}_\pi}\right)^2\frac{1}{(1 + \gamma_\pi^2\theta^2)^2}
\end{equation}
where L is the detector distance. 
Due to the characteristics of the decay kinematics and to the Lorentz boost, if the detector is placed a few degrees off the beam axis, the neutrino energy is no longer proportional to the pion energy but rather has a broad maximum for $\gamma_\pi \theta=1$.
The maximum neutrino energy is $\mathrm{E}_\nu^\star/\theta$, with $\mathrm{E}_\nu^\star=(\mathrm{m}_\pi^2 - \mathrm{m}_\mu^2)/2\mathrm{m}_\pi=29.8 \MeV$ and it depends only from the chosen off-axis angle.
Despite the neutrino yield from any given pion is smaller at an off-axis angle $\theta$ than on-axis, almost all pions in a broad energy range contribute neutrinos in a narrow energy interval around $E_{max}=29.8/\theta \MeV$. 
As a result the neutrino flux is peaked in this narrow energy interval with a flux much larger than that of the on-axis beam at the same energy. This is shown in figure \ref{fig:t2koffspec} for the neutrino spectra at different off-axis angles at 295 km from the source for the T2K long-baseline experiment.

The off-axis technique works well for long-baseline experiments, where the far detector covers a small angular range. 
Here, for a given distance L, the choice of a suitable off-axis angle allows to adjust the beam peak energy E to tune L/E at the maximum of neutrino oscillation. 
Both the currently running long-baseline experiments, T2K \cite{t2k-nim} and NO$\nu$A \cite{NOvA}, are designed with the far detector located off-axis, respectively at $2.4^\circ$ and $0.8^\circ$, tuned to maximise the oscillation probability for their respective distances.
T2K uses Super-Kamiokande as far detector and a neutrino beam produced from protons extracted by a $30 \GeV$ synchrotron at a new accelerator facility (JPARC). 
NO$\nu$A uses a new gigantic plastic scintillator detector located at Ash River, Minnesota, and the NuMI beam in the medium energy configuration, which maximise the off-axis neutrino yield.

For the next generation of proposed long baseline experiments, Hyper-Kamiokande \cite{HyperKamiokande} will also make use of an off-axis beam while the proposed DUNE experiment \cite{DUNE} uses an on-axis beam, aiming to a broader energy spectrum to access both the first and the second oscillation maxima, relying on event by event reconstruction of the neutrino energy in liquid argon.

\subsection{Super beams}

Since the turn of the century, when neutrino oscillation were experimentally established, neutrino physics has focused on precise measurements of all mixing angles and of the squared mass differences, and above all on the search for CP violation in the leptonic sector and establishing the ordering of neutrino masses.
Much of this physics can be accessed by long-baseline experiments tuned at the atmospheric $\Delta m^2$. 
The measurements require very high intensity beams, and a precise knowledge of energy spectra and fractions of the neutrino species ($\Pnumu$, $\APnumu$, $\Pnue$, $\APnue$ ) present in the beam. 
This challenge has stirred interest in the search for new kind of neutrino beams which would overcome limitations and shortcomings of conventional beams.
There is a broad scientific literature, generically going under the tag \textit{Neutrino Factories}, which investigates the feasibility and phenomenology of using different sources for accelerator neutrino beams, like muons (Muon Storage Rings) and radioactive nuclei (Beta Beams), rather than the pions used in conventional beams.
In the next chapter we will discuss these alternative proposals, together with the idea of tagging the flavour of a conventional neutrino beam (Tagged Beams).

In parallel with these new ideas, much technological effort went into the development of conventional beams, with improvements of the focusing devices and of the target, and above all by pushing on the accelerator performance. 
A discussion on the merits of high intensity conventional beams appears already in 2000 in the paper: \textit{Conventional beams or neutrino factories: the next generation of accelerator based neutrino experiments} \cite{Richter2000}.
In that paper, referring to his conjectures on the performance of conventional beam, Richter states: \textit{ "It is well worth the time of the experts to see if my assumptions on potential beam intensity and purity, and background rejection are reasonable. If they are, these experiments can be carried out sooner, and at less cost than those with a muon storage ring source"}. 

Actually, while neutrino factories are still in consideration, conventional horn focused beam of high intensity, the so-called \textit{super beams} are the choice of the present and next generation of long baseline experiments. 
Experiments presently running at JPARC and Fermilab are designed to use high intensity proton beams of about 700~kW.
This is about two order of magnitude higher than that of the K2K long baseline experiment in Japan, the first to confirm at an accelerator beam the oscillation of atmospheric neutrinos \cite{k2k2006}. It is expected that by the end of the next decade a new generation of long baseline experiments will be running with proton beam power well above 1~MW \cite{Derw2012,Iga2016}.

\section{Alternative proposals}
\label{sec:future}
\subsection{Present scenario}
\label{ssec:alternative}
With the conventional techniques of the super beams to be used at least until 2030, there is nowadays no definite plan for the construction of new facilities based on alternative beams and they seem at least deferred to a vary long term strategy as it is suggested in a recent survey of the ICFA Neutrino Panel \cite{Cao2017}: \textit{"The focus of the long-baseline neutrino community has recently been on establishing DUNE and proposing Hyper-K. 
If the science demands a further program with a performance that substantially exceeds that of the ambitious DUNE and Hyper-K experiments, new accelerator and/or detector technologies will be required. 
An R\&D program will be needed to deliver feasible options at the appropriate time.
This R\&D is likely to take many years and needs to be well justified and carefully planned"}.

After the next generation of oscillation experiments, based on high intensity super beams it is likely that further progresses will require breakthrough ideas for innovative and better quality neutrino beams.

Though none of the techniques proposed to replace or complement the conventional beams has come to be a definite project, we feel interesting to discuss in this chapter the more promising approaches still under consideration: Muon Storage Rings, Beta Beams and Tagged Beams. 
A more detailed review of the different technologies proposed for the neutrino factories is given in \cite{Mezz2011}. 

\subsection{Muon storage rings}
\label{ssec:storage}
\begin{figure}[ht]
\centering
\includegraphics[width=0.4\textwidth]{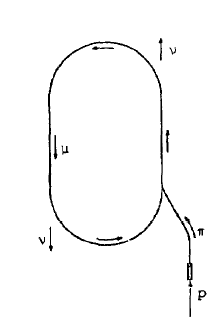}
\caption{ Muon storage ring for neutrino beams \cite{Neuffer81}}
\label{fig:muring} 
\end{figure}

In 1980, David Neuffer proposed to produce neutrino beams from the decay of muons circulating in a dedicated storage ring \cite{Neuffer80,Cline1980,Neuffer81}.
Figure~\ref{fig:muring} from Neuffer's proposal~\cite{Neuffer81} sketches a possible layout of the ring. 
In this scheme, protons from an accelerator hit a target, producing pions. 
Bending magnets drive pions in a large momentum interval into the straight section of the ring, where they decay into muons.
Muons in the correct momentum interval (e.g. $8.0 \pm 0.3$ GeV in one of Neuffer's proposal) and angular acceptance are captured in the ring. 

Owing to the large difference in lifetime between the pion and the muon ($\tau_{\pi} = 2.6 \times 10^{-8}$ s, $\tau_{\mu} = 2.2 \times 10^{-6}$ s), after the fast decay of the pions, only muons continue to circulate in the ring. During their lifetime they produce collimated neutrino beams along the two straight sections.
Since the characteristics of the $ \mu^- \rightarrow \mathrm{e^-} \APnue \Pnumu $ decay (or charge conjugate) are perfectly known, by monitoring the muon beam in the ring it is possible to precisely monitor intensity, composition and energy spectrum of the resulting neutrino beam. This, together with the unique feature of a beam of equal intensity of $\APnue$ and $\Pnumu$ (or of $\Pnue$ and $\APnumu$), makes the muon storage ring a very attractive option for the study of neutrino oscillation. 

Yet, in its simplest implementation the efficiency for collecting muons in the storage ring is too low to produce beams competitive with the usual beams from pions and kaons, and the idea had no follow-up until the discovery of neutrino oscillation. 
In the late 1990's, with the need for high precision measurements of the neutrino mixing matrix, the idea of neutrinos from muons received new attention and evolved in the design of the so-called Neutrino Factory. 
To create neutrino beams of high intensity, a Neutrino Factory would make use of an intense proton beam to produce a beam of low energy pions. The muons produced in the following pion decay are then phase space compressed (\textit{cooled}). 
After the cooling, muons are accelerated again to the desired energy, before to be finally injected into a storage ring with long straight sections pointing in the desired directions. 
\begin{figure}[ht]
\centering
\includegraphics[width=0.7\textwidth]{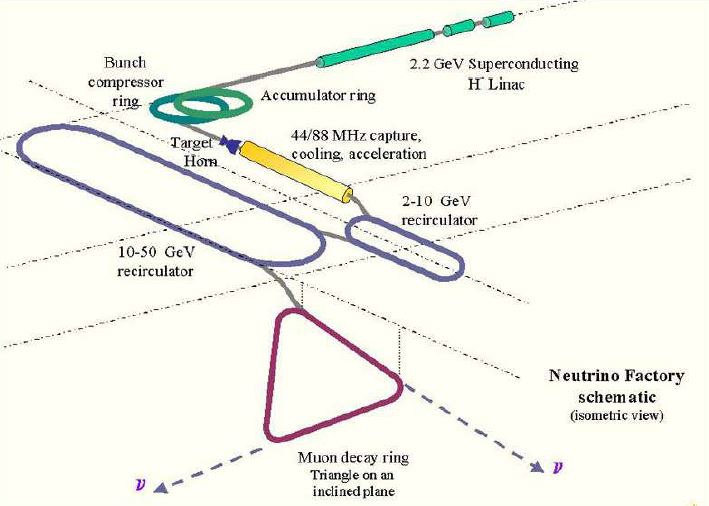}
\caption{ Schematic layout of a Neutrino Factory \cite{Blondel2005} }
\label{fig:nufact} 
\end{figure}
Complex schemes have been studied, with the final storage in a triangular ring, so to have straight sections pointing at detectors at two or three different baselines, like in the schematic layout shown in figure \ref{fig:nufact}.

A schematic report on Neutrino Factory projects is given in \cite{Blondel2005}, where costs between $1\mbox{--}2$B US dollars are quoted.
In spite of these high costs, much R\&D has continued to be devoted to the subject, also because the production of intense monochromatic muon beams would serve to the construction of a muon collider (e.g. a Higgs factory). 
A more recent review of the status of the studies on Neutrino Factories can be found in \cite{Mezz2011}. 

However, nowadays Neutrino Factories appear no longer competitive with respect to the planned super beams, which promise to enable detailed studies of neutrino oscillation, including access to a broad range of possible values of the CP violating phase. 
An important reason for this is that the $\theta_{13}$ mixing angle regulating the $\Pnumu - \Pnue$ oscillation at the atmospheric $\Delta m^2$ is relatively large, easing the search for CP violations. 
It is also important to note that in Neutrino Factories, the advantage of having available the flux of $\Pnue$ or $\APnue$ accompanying the $\APnumu$ or $\Pnumu$ is fully exploitable for oscillation studies, only if the gigantic neutrino detectors are equipped for the measurement of the charge of the muon produced in neutrino interactions.

By now, as described in \cite{Long2018}, the only advanced project is nuSTORM, a first level proposal to the Fermilab PAC \cite{Adey2013}. 
The nuSTORM muon storage ring is very similar to that proposed by Neuffer in 1980 and does not foresee the muon cooling and acceleration, which results in a reduced neutrino flux.
The proponents envisage using 120~GeV protons to produce pions off a solid target. 
The pions are collected by magnetic fields (horn and quadrupoles) and injected in the straight section of the ring.
Muons from pion decay are stored in the ring if they have a momentum of $3.8 \GeV\pm 10\%$.
Neutrino beams are produced by the muon decays in the 226~m long straight sections. The physics goals are the search for oscillation into a sterile neutrino and the measurement of low energy neutrino cross-sections.

\subsection{Beta beams}

The name {\it beta-beam} refers to the production of a pure beam of electron neutrinos or antineutrinos through the beta decay of accelerated radioactive ions circulating in a storage ring \cite{betabeam}. 
In the original conceptual proposal, using established techniques (e.g those of the ISOLDE facility at CERN \cite{IONS}), ions produced by an intense radioactive source are accelerated to energies, about 150~GeV/nucleon, similar to the ones achieved at the heavy ion programme at CERN PS/SPS. 
Such a high energy radioactive beam, injected in a storage ring would be the source of a pure electron neutrino beam (or antineutrino, depending on the accelerated ion).
The neutrino transverse momentum with respect to the beam axis is equal to the neutrino transverse momentum in the ion rest frame while the longitudinal momentum is boosted by the Lorentz $\gamma$ of the accelerated ion and therefore the typical neutrino emission angle goes like $1/\gamma$. 
The use of a radioactive nucleus with a small $\mathrm{Q}$ value improves the neutrino beam collimation and the neutrinos yield along the axis of the storage ring straight section. 
Unfortunately the ion lifetime is inversely proportional to $\mathrm{Q}^5$ and the choice between candidate ions must be carefully optimised. 
${}^{18}\mathrm{Ne}$ and ${}^6\mathrm{He}$ where identified as the best sources of respectively electron neutrinos and antineutrinos due to their small $\mathrm{Q}$ values, about $3.5 \MeV$, and their lifetime, of the order of 1s, which are not too long to require a proportionally higher number of ions circulating in the storage ring for a given neutrino flux. 

In conventional muon neutrino beams the contamination from different neutrino species is inevitable and the knowledge of the neutrino spectrum and flux involves sizeable systematic uncertainties. 
In contrast beta beams are pure electron (anti)neutrino beams and the energy spectrum and flux can be easily calculated from the number of circulating ions, their known decay kinematic and Lorentz boost. 
The neutrino energy is $\mathrm{E}_\nu = 2\gamma E_\nu^\ast$, where $E_\nu^\ast$ is the neutrino energy in the ion rest frame.
For ${}^6\mathrm{He}$ ions accelerated at $\gamma=150$, the average neutrino energy would be $580 \MeV$.
Accelerated under the same conditions, ${}^{18}\mathrm{Ne}$ ions would reach $\gamma=250$ due to the larger Z/A and the neutrino energy would be $930 \MeV$, despite the two ions have a similar decay spectrum.
It is remarkable that at a beta beam facility an experiment with muon identification capability (even without charge identification) would be able to search for CP violation by comparing the probabilities $\Pnue\rightarrow\Pnumu$ and $\APnue\rightarrow\APnumu$.

While the conventional neutrino beam is a tertiary beam, obtained from the decay of secondary mesons produced from accelerated primary protons, neutrinos from both the beta beam and the neutrino factory are produced directly from the decay of their accelerated parents.

The beta beam idea stirred interest in both the accelerator physics and neutrino phenomenology communities and several studies were realised in parallel with the neutrino factory studies when the goal appeared to be the search for a possibly vanishing small $\theta_{13}$. 
Ions acceleration at $\gamma~100$, corresponding to neutrino energies of a few hundreds of $\MeV$ and calling for experimental baselines of the order of 100 km, were considered \cite{Mezzetto2003} as well as higher energy options, up to $\gamma~2000$ with baselines of 2000 Km \cite{Burguet2004} and also a low energy option, down to neutrinos of tens of $\MeV$, to measure cross-sections relevant for nucleosynthesis and supernova explosions \cite{Volpe2004}.

The complexity and cost of beta beams and neutrino factories, and finally the discovery that $\theta_{13}$ is almost as large as it could have been, have shifted the interest back to conventional neutrino beams for the next generation of 
experiments in the quest for leptonic CP violation. 

\subsection{Tagged beams}

Experiments with neutrino beams would greatly profit from a precise knowledge of the flavour and the energy of the neutrinos, and even more if that information is unambiguously associated, event by event, to the neutrino interactions in the detector. 
Flavour knowledge is clearly extremely important for oscillation studies. 
The addition of the energy information would allow to improve the study of neutrino cross-sections, which nowadays are still affected by large uncertainties, especially in the 1~GeV energy region. 
The ideal technique would be a tagged neutrino facility, where a precise knowledge of the flavour of the neutrino and possibly a determination of its energy, are achieved by measuring, in the decay of the meson, the lepton associated to the neutrino. 
More information is obtained if the tagging detector at the decay is able to work in time coincidence with the interaction in a downstream neutrino detector.
 
Almost forty years ago Bruno Pontecorvo wrote in his typical understated style {\it ''the possibility of using tagged-neutrino beams in high-energy experiments must have occurred to many people''} \cite{Pontecorvo1979}.
In order to become reality this conceptually simple and elegant idea has to face big experimental challenges. 
To apply the idea of the tagging to a conventional accelerator neutrino beam, the main challenge is that the neutrino source is a meson decay taking place in a tunnel which span several decay lengths, typically tens or hundreds of meters, in a harsh, high rate, high radiation environment. 
One of the oldest idea \cite{Hand1969} is based on the detection of the muons to tag $\Pnumu$ from $\PK_{\mu 2}$ decays, taking advantage of the large difference in Q value between pion and kaon decays. 
A different approach \cite{Bernstein1988}, consists in tagging neutrinos from a $\PK\mathrm{_L}$ beam, by measuring the out going charged pion and lepton at the decay vertex $\PK\mathrm{_L}\rightarrow \pi \mathrm{e} \nu_\mathrm{e}$ and $\PK\mathrm{_L}\rightarrow \pi \mu \nu_\mu$. 
Measuring the positrons (electrons) in $\mathrm{K}_{\mathrm{e}3}$ decays in the decay tunnel of a conventional neutrino beam, was proposed \cite{antitag} to tag and veto the prompt electron (anti)neutrino contamination in experiment searching for the $\Pnumu\rightarrow\Pnue$ oscillation.

In recent years, the relevance of a precise knowledge of neutrino cross-sections, in particular electron neutrino cross-sections, for the next generation of long baseline experiments searching for leptonic CP violation, has driven the proposal of a dedicated facility for $\Pnue$ and $\APnue$ beams.
This facility \cite{LLT} is a narrow band beam where the flux uncertainty, which is the main limitation for a precise cross-section measurement, would be reduced at one percent level by directly measuring the positrons (electrons) from $\PK_{\mathrm{e}3}$ decays in a large volume calorimeter surrounding the decay tunnel. 
ENUBET\footnotemark, a dedicated detector R\&D, is under way to prove the technical feasibility of the proposed design.

Several decades after Pontecorvo we are in all probability still not so close to see tagged neutrino beams. Nevertheless we like to quote his optimistic view that {\it "in spite of the difficulty it seems that sooner or later such facilities will be available at various high-energy accelerators"} \cite{Pontecorvo1979}.

\footnotetext{Funded by EU Horizon-2020 Research and Innovation programme, GRANT n.681647}

\section{Conclusions}
\label{conclusions}

The two-neutrino experiment in 1962 resulted not only in a fundamental discovery, it was also the first demonstration of a new tool, neutrino beams produced at a proton accelerator, which in more than half a century has driven major advances in our understanding of nature.

A fundamental role for these advances has been played by the intensity of the neutrino beams, and before to conclude, we have to remark that, together with the new ideas and improvements specifically developed for the neutrino beams, at the origin of the increase of neutrino beam intensity is the astonishing development in the domain of proton accelerators.

\begin{figure}[hbt]
\centering
\includegraphics[width=0.75\textwidth]{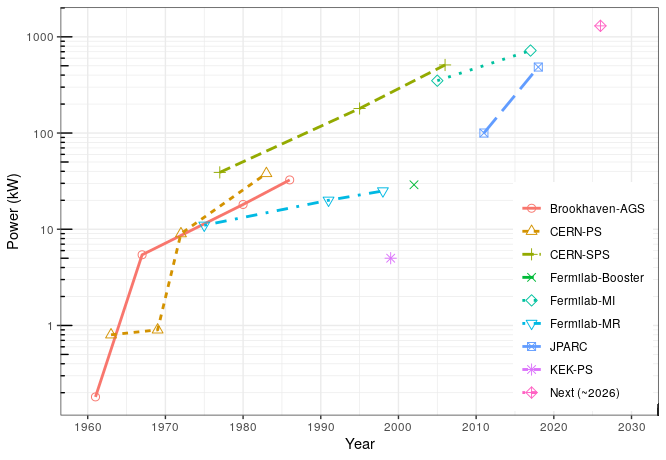}
\caption{Proton intensity for neutrino beam lines, from the pioneering beams of the 1960's to the future generation of long-baseline oscillation experiments before the end of the next decade.}
\label{fig:bpower} 
\end{figure}

Figure~\ref{fig:bpower} shows the beam power record intensities at the proton accelerators used to produce neutrino beams in some of the major laboratories. 
Compared to the few hundreds watts of power of the first neutrino beams, we have several hundreds kilowatts today and up to $1\mbox{--}2 \mathrm{MW}$, projected for the next generation of long baseline experiments.

Finally, to compare the development in beam intensity and detector masses, consider that in the 1970's, the discovery of the neutral current interaction with Gargamelle was made with a detector of $5 \mathrm{t}$ fiducial mass, exposed to $10^{17}$ protons on target.
The presently running long-baseline neutrino oscillation experiments, T2K and NO$\nu$A, have collected in excess of $10^{21}$ protons on target, with detector masses about four orders of magnitude larger than Gargamelle.

Accelerator neutrino beams of extremely high intensity are fundamental for the next generation of experiments, holding promise for the unveiling of the secret of leptonic CP violation and for challenging with unprecedented precision our understanding of neutrino properties.

\end{document}